\documentclass{aa}  

\usepackage{graphicx}
\usepackage{txfonts}
\usepackage{siunitx}
\DeclareSIUnit\gauss{G}
\usepackage{ragged2e}
\usepackage{natbib}
\usepackage{subfigure}
\usepackage{hyperref}
\hypersetup{
    colorlinks=true,
    linkcolor=blue,
    citecolor=blue,
    filecolor=magenta,      
    urlcolor=cyan,
}

\newcommand{\va}{v_{\rm A}}

\begin{document}

   \title{Alfvén wave heating in partially ionized thin threads of solar prominences}

   \author{Lloren\c{c} Melis
          \inst{1,2}
          \and
          Roberto Soler\inst{1,2}
          \and
          Jos\'e Luis Ballester\inst{1,2}
          }

   \institute{Departament de Física, Universitat de les Illes Balears, E-07122, Palma de Mallorca, Spain\\ \and Insitut d'Aplicacions Computacionals de Codi Comunitari (IAC3), Universitat de les Illes Balears, E-07122, Palma de Mallorca, Spain}

 
  \abstract{There is observational evidence of the presence of small-amplitude transverse magnetohydrodynamic (MHD) waves with a wide range of frequencies in the threads of solar prominences. It is believed that the waves are driven at the photosphere and propagate along the magnetic field lines up to prominences suspended in the corona. The dissipation of MHD wave energy in the partially ionized prominence plasma is a heating mechanism whose relevance needs to be explored. Here we consider a simple 1D model for a non-uniform  thin thread and investigate the heating associated with  dissipation of Alfv\'en waves. The model assumes an ad hoc density profile and a uniform pressure, while the temperature and ionization degree are self-consistently computed  considering  either  LTE  or  non-LTE  approximations for the hydrogen ionization. A broadband driver for Alfv\'en waves is placed at one end of the magnetic field line, representing photospheric excitation. The  Alfv\'enic perturbations  along the thread are obtained by solving the linearized  MHD  equations  for  a  partially  ionized  plasma  in  the single-fluid  approximation. We find that wave heating in the partially ionized part of the thread is significant enough to compensate for energy losses due to radiative cooling. A greater amount of heating is found in the LTE case because the ionization degree for core prominence temperatures is  lower than that in the non-LTE approximation. This results in a greater level of dissipation due to ambipolar diffusion in the LTE case. Conversely, in the hot coronal part of the model, the plasma is fully ionized and wave heating is negligible. The results of this simple model suggest that MHD wave heating can be relevant for the energy balance in prominences. Further studies based on more elaborate models are required.}


   \maketitle

\section{Introduction}

Solar prominences are fascinating inhabitants of the solar corona that are made of relatively cool and dense plasma whose physical properties are akin to those in the chromosphere \citep[see, e.g.,][]{vialengvold2015}. High-resolution observations show that prominences are formed by a myriad of thin and long sub-structures, usually called threads, which are believed to outline particular field lines of the prominence global magnetic structure that is anchored at the photosphere \citep[see, e.g.,][]{lin2011,martin2015}. While it has been well established that the magnetic field provides the necessary force to support the prominence material against gravity, the energy balance in the prominence plasma is less understood. This balance requires an understanding of the processes involved with heating and cooling \citep[see, e.g.,][]{parenti2014,gilbert2015,heinzel2015book}. According to \citet{gilbert2015}, radiative heating due to the illumination from the surrounding solar atmosphere may be the dominant heating mechanism, however, determining the radiation field within prominences is not an easy task \citep{heinzel2015book}. In addition, the prominence models that are obtained when balancing incident radiation and cooling \citep[e.g.,][]{Heasley1976,Anzer1999,Heinzel2010,Heinzel2012} suggest that an additional source of heating must be present to explain the temperatures at prominence cores, which are estimated to be around 6000--8000 \si{\kelvin} \citep[see][]{parenti2014}. Models of prominence threads under mechanical and thermal equilibrium typically require ad hoc heating to reproduce the expected density and temperature profiles \citep[see the recent work by][]{terradas2020}. So, the problem of energy balance remains open.

Observations show the ubiquitous presence of waves and oscillations in the fine structure of prominences \citep[see the review by][]{arregui2018}, which are interpreted as magnetohydrodynamic (MHD) waves \citep{ballester2015}. \citet{hillier2013} suggested that transverse MHD waves in prominence threads are driven by horizontal motions in the photosphere. In this scenario, MHD waves can transport energy from the photosphere up to prominences suspended  in the corona \citep[see][]{soler2019}. If an efficient enough dissipation mechanism is at work in the prominence plasma, MHD wave energy could be thermalized. Therefore, wave heating arises as another viable heating mechanism in prominences whose actual relevance needs to be determined. 

The prospects of wave heating in prominences were previously discussed by \citet{pecseli2000} and \citet{2007A&A...469.1109P}, but their conclusions were not in agreement. \citet{pecseli2000} estimated that the heating associated with the  dissipation of Alfv\'en waves due to ion-neutral collisions would only contribute a tiny fraction to the bulk radiative energy. However, \citet{pecseli2000} only considered Alfv\'en waves with a fixed frequency, while observational evidence has pointed to the presence of a broadband spectrum \citep{hillier2013}.  Conversely, \citet{2007A&A...469.1109P} estimated the Alfv\'en wave energy flux from measured nonthermal velocities and suggested that a large fraction of radiative losses could be compensated for. In their estimation, \citet{2007A&A...469.1109P} assumed that the entire energy flux would be dissipated, while the actual dissipated energy should depend on the efficiency of the damping mechanism at work. 

To our knowledge, the first detailed investigation of wave heating in prominences was performed in \citet{soler2016} using a very idealized configuration. The prominence was represented by a homogenous slab with a constant transverse magnetic field embedded in a homogeneous corona. No fine structure was considered. The prominence plasma was assumed to be partially ionized with an arbitrarily prescribed ionization degree, while the action of ion-neutral collisions  was adopted as a dissipation mechanism. Alfv\'en waves were launched towards the prominence slab and the plasma heating rate due to wave dissipation was computed. \citet{soler2016} estimated the volumetric heating integrated over the range of periods between 0.1 s and 100 s to be as large as 10\% of the bulk radiative energy of the cool prominence plasma. Despite the important idealizations of the model of \citet{soler2016}, their results provide evidence supporting the potential of Alfv\'en wave heating in prominences. 

The purpose here is to go further and deeper into the problem of prominence wave heating by improving the treatment of \citet{soler2016}. Although it is still simple, we shall consider a more elaborated model for a 1D thin thread with a  longitudinal magnetic field. After assuming a constant pressure along the thread and imposing an ad hoc density profile, the temperature, and ionization profiles will be self-consistently computed considering both LTE and non-LTE approximations for the hydrogen ionization degree. The considered dissipation mechanisms will be Ohm's diffusion due to magnetic resistivity and ambipolar diffusion due to ion-neutral collisions. A broadband driver for Alfv\'en waves will be considered at one end of the magnetic field line,  aiming to mimic photospheric excitation. The plasma heating rate due to the dissipation of the waves will be computed and their relevance for the energy balance will be discussed.

This paper is organized as follows. Section~\ref{sec:back} contains a description of the thin thread model. Section~\ref{sec:method} includes the basic equations for the discussion of Alfv\'en waves and a description of the numerical method. The results are presented and discussed in Section~\ref{sec:results}, while some concluding remarks and prospects for future works are given in Section~\ref{sec:conc}.

\section{Background configuration}
\label{sec:back}

\subsection{Thin thread model}
\label{subsection:model}
\begin{figure*}
    \centering
    \includegraphics[width=6cm,height=4cm]{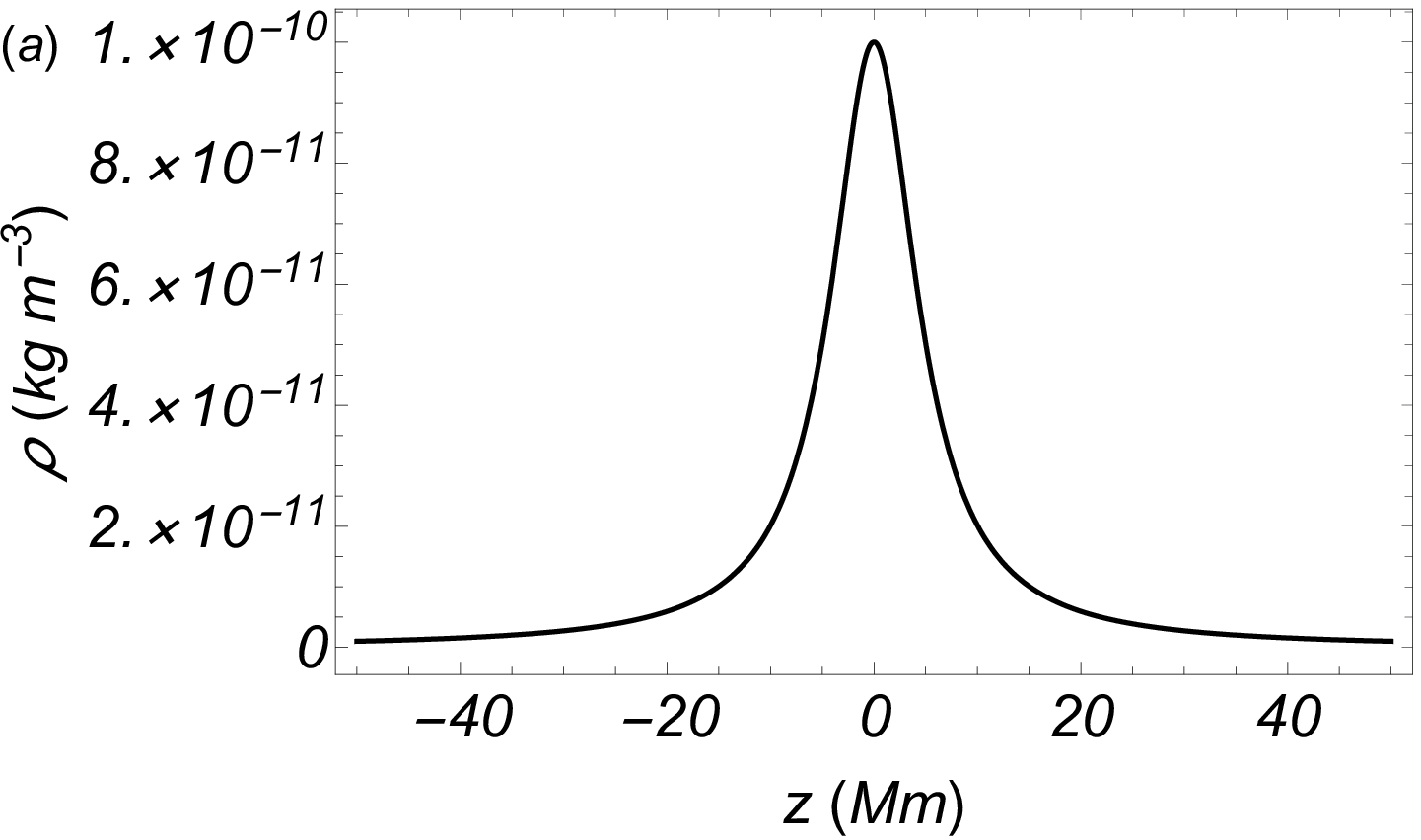} 
    \includegraphics[width=6cm,height=4cm]{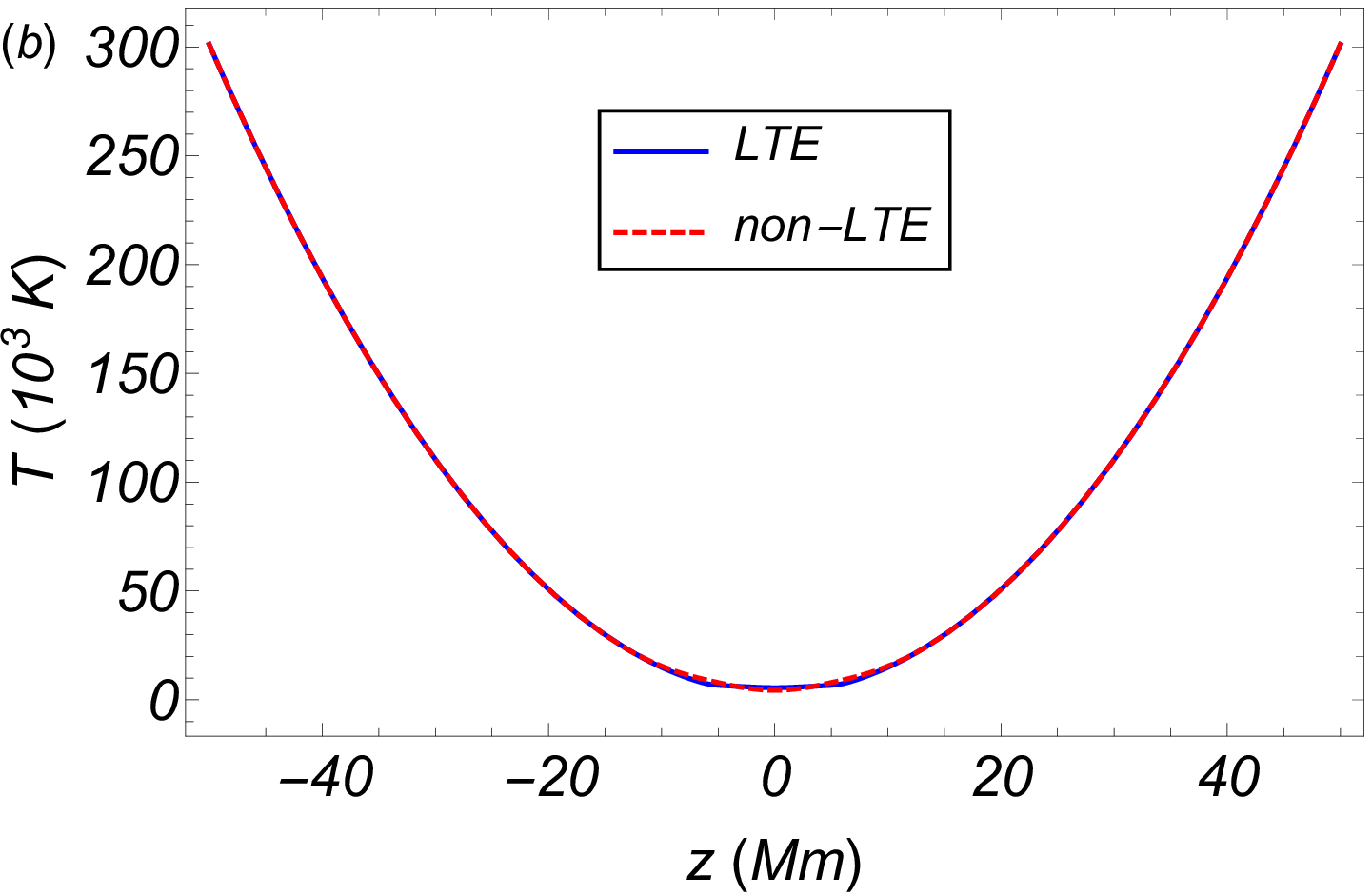} 
    \includegraphics[width=6cm,height=4cm]{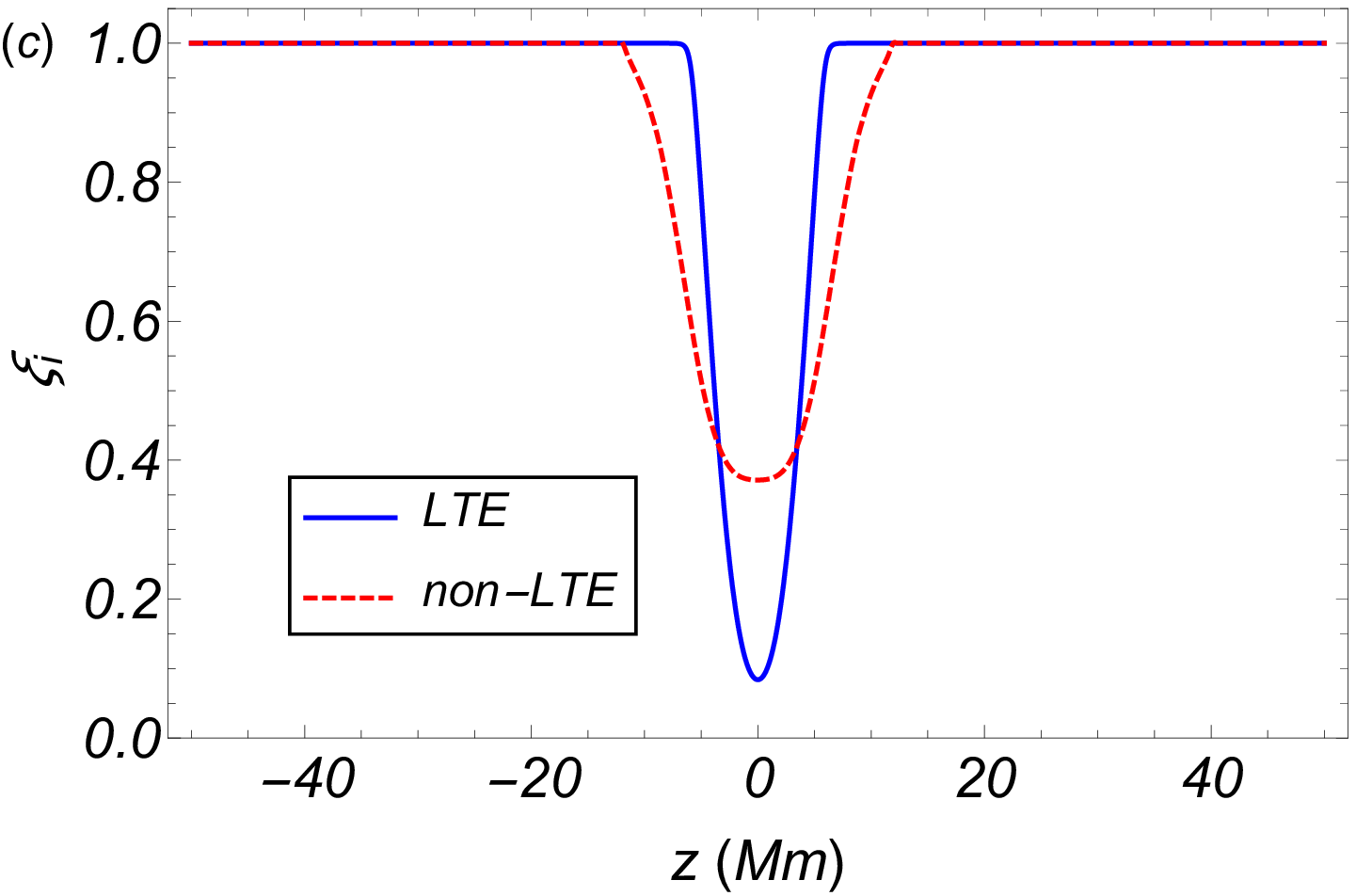} 
    \includegraphics[width=6cm,height=4cm]{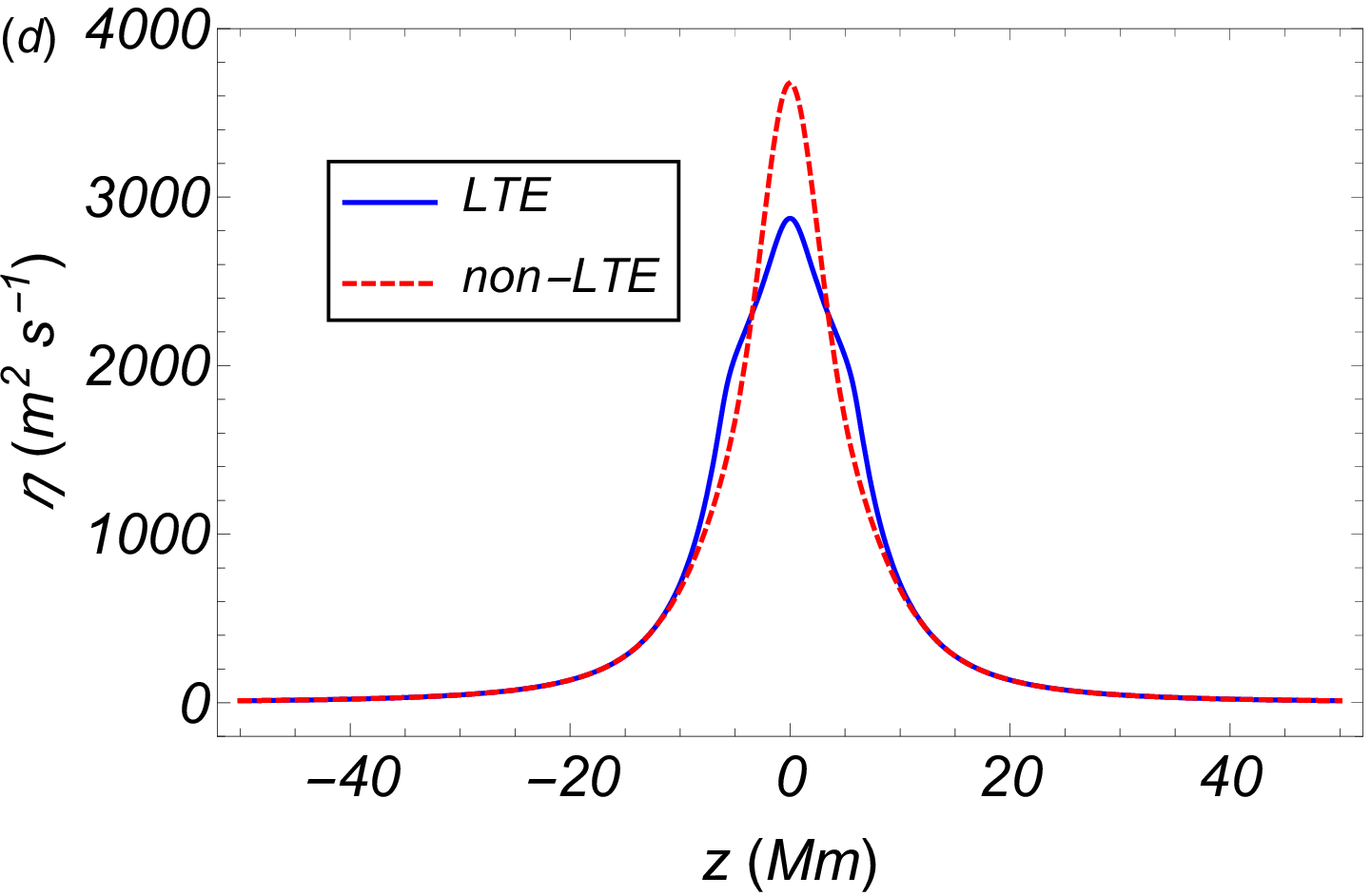} 
    \includegraphics[width=6cm,height=4cm]{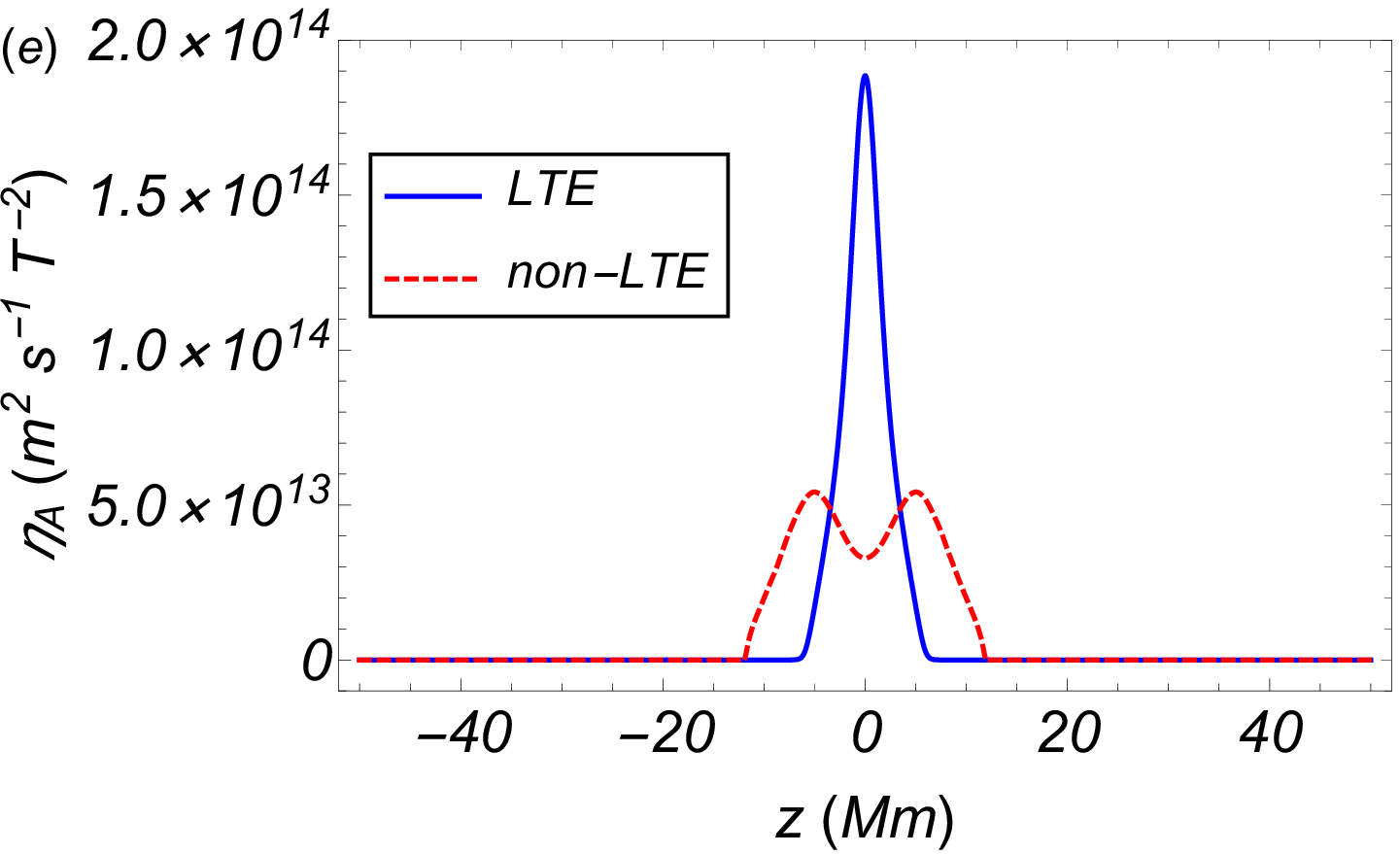} 
    \includegraphics[width=6cm,height=4cm]{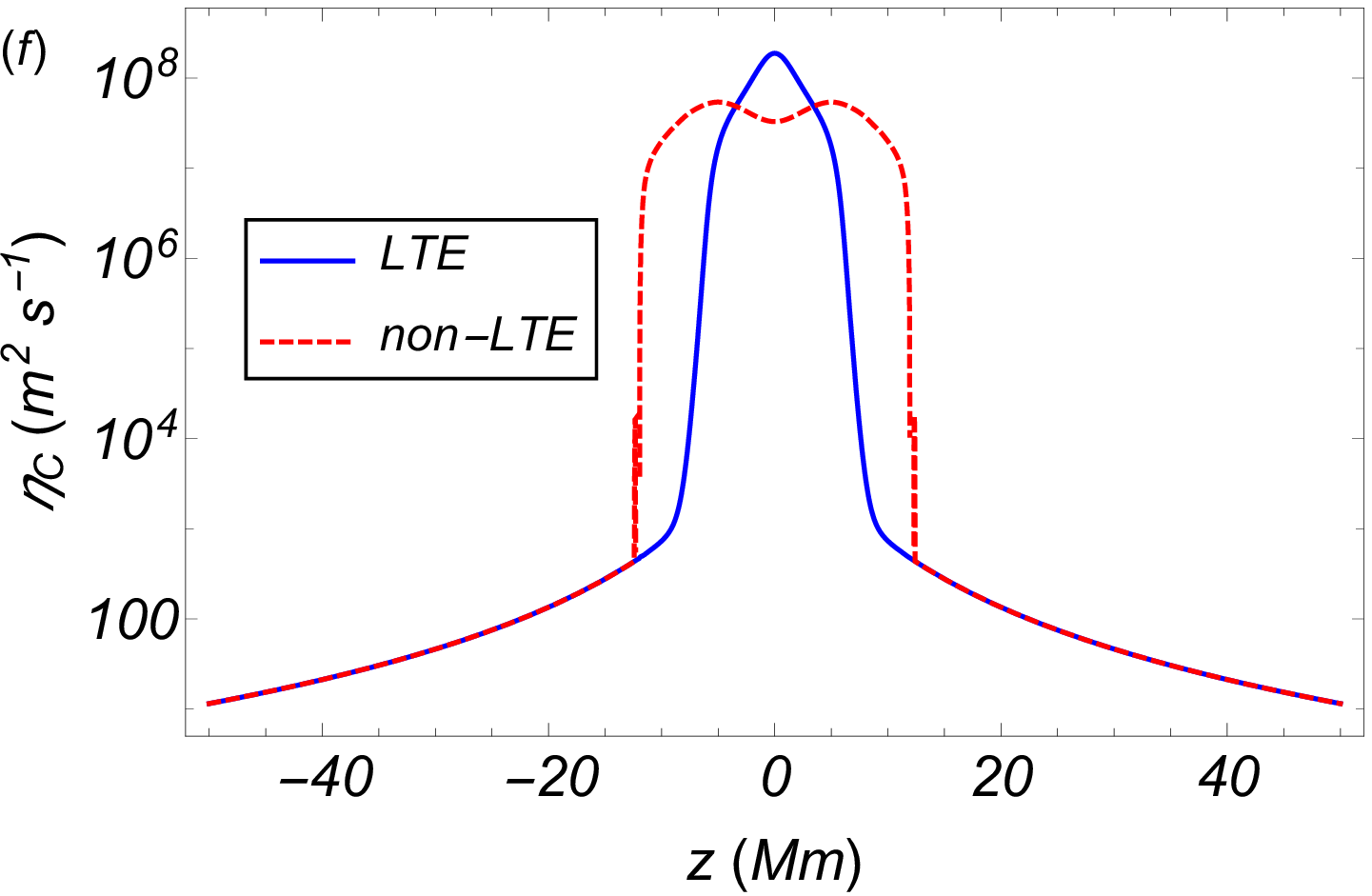} 
    \caption{Background magnitudes along the thread; \textbf{a)} Adopted density profile; \textbf{b)} Computed temperature; \textbf{c)} Computed ionization fraction; \textbf{d)} Ohm's diffusion coefficient; \textbf{e)} Ambipolar diffusion coefficient; \textbf{f)} Cowling's diffusion coefficient. Results for the LTE and non-LTE methods in Figures \textbf{b)} to \textbf{f)}.}
    \label{fig:magnitudes}
\end{figure*}

We consider a simple 1D model for a prominence thin thread  made of a single magnetic field line of length $L = 10^8$~m. We neglect curvature and the effect of gravity, so that the magnetic field line is straight and aligned with the $z$-axis for convenience. The constant magnetic field strength is denoted by $B$. A typical value in quiescent prominences is $B=10$~G. The center of the thread is located at $z=0$ and its ends are at $z=\pm L/2$. The pressure and density variation along prominence threads is linked to the thermal instability process that leads to the condensation of cool and dense plasma along the prominence magnetic field lines \citep[see, e.g.,][]{luna2012}. Recently, \citet{terradas2020}  explored models  of 1D prominence threads under hydrostatic and thermal equilibrium. The consideration of such detailed models is beyond the aim and purpose of the present work. Instead, we assume for simplicity that the thread has a constant pressure of $p=5 \times 10^{-3}$ Pa and that the density variation along the thread follows a Lorentzian dependence that aims to mimic the expected density profile along prominence threads \citep[see][]{soler2015} and is a plausible choice for the density profile according to Bayesian model comparison \citep[see][]{arregui2015}, namely:
\begin{equation}
    \rho(z) = \frac{\rho_{0}}{1+4(\chi-1)z^{2}/L^{2}},
\label{eq:density}    
\end{equation}
\justify
where $\rho_{0}$ = $10^{-10}$ \si{\kilogram\per\metre\cubed} is the density at the center of the thread and $\chi=100$ is the ratio between the density at $z=0$ and that at $z=\pm L/2$. The density profile is represented in Figure \ref{fig:magnitudes}a.
Once we have defined the pressure and density of the thread, the next step is obtaining the profiles along $z$ for the temperature, $T$, and the ionization fraction, $\xi_{i}$, which is defined as the ratio of the ion density  to the total density of the plasma. Due to the relatively low temperatures of prominences, the plasma is only partially ionized \citep[see, e.g.,][]{labrosse2010}. Hence, the ionization fraction can vary from  fully ionized plasma, $\xi_{i} = 1$, to almost neutral gas, $\xi_{i} \approx 0$. Both $\xi_{i}$ and $T$ are related through the ideal equation of state, which for a plasma only composed by hydrogen reads
\begin{equation}
p = \left(1 + \xi_{i} \right) \rho R T,    
\end{equation}
where $R$ is the ideal gas constant. However,  we need another independent relation between $\xi_{i}$ and $T$ to close the system. This additional relation is the one that determines the ionization state of the plasma. Computing the ionization degree of a prominence thread in a consistent way requires the full solution of the radiative transfer equations taking into account the incident radiation \citep[see, e.g.,][]{gouttebroze2009}. To avoid that complicated calculations, here we use two different approximate approaches. The first one uses the Saha equation, which assumes local thermodynamic equilibrium (LTE) and the second one uses the tabulated values of   the ionization fraction as a function of the temperature and pressure given by \citet{heinzel2015}, which are based on non-LTE radiative transfer computations in 1D prominence slabs\footnote{We note that in the computations of \citet{heinzel2015}, 10~$\%$ of the plasma was composed of neutral helium, whereas here we ignore helium for the sake of simplicity.}.  One of the purposes of this work is to compare the results obtained for both approaches. Throughout the paper, we use ``LTE case'' to refer to the method that involves the Saha equation and ``non-LTE case'' to refer to the one based on the computations of  \citet{heinzel2015}.

\begin{figure}[htbp]
    \resizebox{\hsize}{!}{\includegraphics{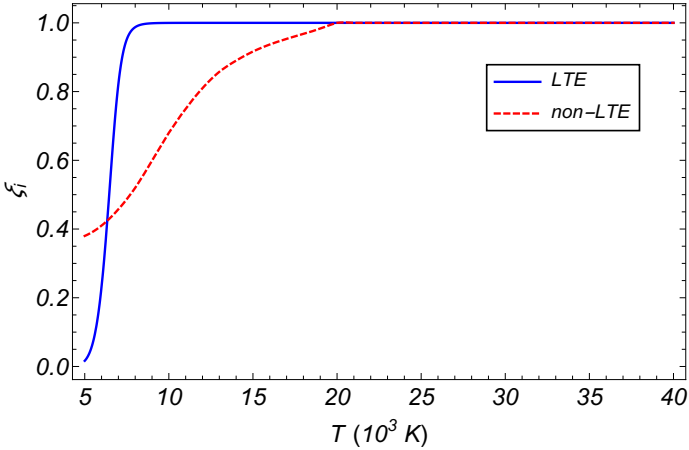}}
    \caption{Ionization fraction, $\xi_{i}$, as a function of temperature, $T$, for the LTE and non-LTE methods.}
    \label{fig:ion}
\end{figure}

On the one hand,  the expression of $\xi_{i}$ according to the Saha equation in the case of a hydrogen plasma is
\begin{equation}
\xi_{i}=\frac{1}{2}\mathcal{M}\left(\sqrt{1+\frac{4}{\mathcal{M}}}-1\right),
\label{eq:saha}    
\end{equation}
where, in MKS units,
\begin{equation}
\mathcal{M} \approx 4 \times 10^{-6} \rho^{-1} T^{3/2} \exp(-T^{*}/T),
\end{equation}
with $T^{*}=1.578 \times 10^{5}$ K. On the other hand, in the case of the numerical results by \citet{heinzel2015} we use the  values given in Table~\ref{tab:heinzel}, which relate the temperature and the ionization fraction for the constant pressure considered here. In Figure \ref{fig:ion}, we plot the ionization fraction as a function of temperature for both LTE and non-LTE approaches. In the case of the non-LTE method, we interpolated the values of Table~\ref{tab:heinzel}  using a third-order polynomial and have imposed $\xi_{i}=1$, that is, full ionization, for $T \geq$~20,000~K. Extrapolation has been used when $T < $~6,000~K.  When the temperature increases, the ionization fraction in the LTE case reaches the completely ionized state much faster than in the case of the non-LTE case.

\begin{table}[!tb]
    \centering
    \begin{tabular}{|c|c|c|c|c|c|}
    \hline
     \textbf{$T$} & 6000 & 8000 & 10000 & 12000 & 14000 \\ \hline
     \textbf{$\xi_{i}$} & 0.41 & 0.52 & 0.68 & 0.81 & 0.89 \\ \hline 
    \end{tabular}
    \caption{Values of the ionization fraction, $\xi_{i}$, for various values of the temperature (in K) for a gas pressure of $5 \times 10^{-3}$~Pa and an altitude above the photosphere of 20,000~km. Adapted from \citet{heinzel2015}.}
    \label{tab:heinzel}
\end{table}

The resulting profiles of temperature and ionization fraction along the thread computed for both approaches can be seen in Figures~\ref{fig:magnitudes}b and \ref{fig:magnitudes}c, respectively. For the temperature, both methods give a similar result: a slightly modified parabolic-like profile with the minimum at the center of the thread and the maximum at the ends. The central temperature is $\sim$~5000~\si{\kelvin} and the temperature at the ends is $\sim$~\SI{3e5}{\kelvin}. Such values are consistent with what is expected in prominences. For the ionization fraction, both approaches provide a partially ionized zone around the thread center, where the temperature is lowest and the density is largest. The LTE method gives a narrower partially ionized region and the minimum value of $\xi_{i}$ is less than half that obtained in the non-LTE method, for which there is a wider partially ionized region but the minimum $\xi_{i}$ is larger. The ionization fraction at the center is $\sim$~0.1 for the LTE case and $\sim$~0.37 for the non-LTE case.

\subsection{Diffusion coefficients}

In partially ionized plasma, Ohm's and ambipolar diffusion are two mechanisms that can dissipate Alfv\'en waves \citep[see, e.g.,][]{ballester2018}. Ohm's diffusion (also called electrical resistivity) is caused by the collisions of electrons with ions and neutrals, while ambipolar diffusion is caused by the collisions of neutrals with ions and electrons. The Ohm's, $\eta$, and ambipolar, $\eta_{A}$, coefficients are calculated as:
\begin{eqnarray}
        \eta &=& \frac{\alpha_e}{\mu_{0}e^2 n_e^2}, \label{eq:ohm}  \\
            \eta_{\rm A} &=& \frac{\xi_{n}^{2}}{\mu_{0}\alpha_{n}}, \label{eq:ambipolar}    
\end{eqnarray}
where $\mu_{0}$ is vacuum magnetic permeability,  $n_{e}$ and $e$ are the electron number density and charge, respectively, $\xi_{n}=1-\xi_{i}$ is the neutral atoms fraction, and $\alpha_e$ and $\alpha_n$ are the electron and neutral friction coefficients, respectively. The friction coefficients are computed as:
\begin{equation}
\alpha_\beta = \sum_{\beta' \neq \beta} \alpha_{\beta \beta'},    
\end{equation}
where $\beta$ and $\beta'$ denote either electrons (e), ions (i), or neutrals (n), and $\alpha_{\beta \beta'} = \alpha_{\beta' \beta}$  is the symmetric friction coefficient for collisions between two particular species \citep[see, e.g.,][]{1965RvPP....1..205B}. For collisions between two charged particles, the friction coefficient is: 
\begin{equation}
    \alpha_{\beta \beta'}=\frac{n_{\beta}n_{\beta'}e^{4}\ln{\Lambda_{\beta \beta'}}}{6\pi\sqrt{2\pi}\epsilon^{2}_{0}m_{\beta \beta'}\left(k_{B}T/m_{\beta \beta'}\right)^{3/2}},
\end{equation}
where $m_{\beta \beta'} = m_\beta m_\beta' / \left( m_\beta + m_\beta' \right)$ is the reduced mass, $e$ is the electron charge, $k_{B}$ is Boltzmann's constant, $\epsilon_{0}$ is the permittivity of free space, $n_{\beta}$ is the number density, and $\ln \Lambda_{\beta \beta'}$ is Coulomb's logarithm, given by \citep[see, e.g.,][]{vranjes2013}:
\begin{equation}
    \ln{\Lambda_{\beta \beta'}}=\ln{\left(\frac{24\pi\epsilon_{0}^{3/2}k_{B}^{3/2}T^{3/2}}{e^{3}\sqrt{n_{\beta}+n_{\beta'}}}\right)}.
\end{equation}
If at least one of the colliding particles is a neutral, so that $\beta'=n$, the friction coefficient is
\begin{equation}
    \alpha_{\beta n}=n_{\beta}n_{n}m_{\beta n}\left[ \frac{8k_{B}T}{\pi m_{\beta n}} \right]^{1/2} \sigma_{\beta n},
\end{equation}
where $\sigma_{\beta n}$ is the collision cross-section.

The profiles of the diffusion coefficients are presented in Figure \ref{fig:magnitudes}d for Ohm's diffusion and in Figure~\ref{fig:magnitudes}e for  ambipolar diffusion. The profiles of the Ohm's coefficient are similar in the two cases.  The Ohm's coefficient for the LTE case has a lower maximum but its profile is slightly wider than in the non-LTE case. In both cases, the maximum value is situated at the center of the thread and is $\sim$~\SI{3e3}{\metre\per\second\squared} for the LTE case and $\sim$~\SI{4e3}{\metre\per\second\squared} for the non-LTE case. Conversely, the profiles of the ambipolar coefficient are significantly different in the LTE and non-LTE cases. The ambipolar coefficient for the LTE case takes the maximum value, $\sim$~\SI{1.6e14}{\metre\per\second\squared\per\tesla\squared},  at the center of the thread, whereas for the non-LTE case the maximum, $\sim$~\SI{5.4e13}{\metre\per\second\squared\per\tesla\squared}, is displaced from the center and instead of a maximum there is a relative minimum at the center of the thread. Comparing the relative numerical values of both diffusion coefficients  and assuming typical values of the magnetic field  strength in prominences, it turns out that the effect of ambipolar diffusion should be much larger than that of Ohm's diffusion, so that the ambipolar term is expected to be dominant in the calculation of wave dissipation. This simple estimation agrees with previous results from, for instance, \citet{2011A&A...525A..60B} regarding the efficiency of Ohm's and ambipolar diffusion in prominence conditions.

\section{Method}
\label{sec:method}

\subsection{Basic equations}

To study the propagation of Alfv\'en waves along the thread, we used the MHD equations for a partially ionized plasma under the single-fluid approximation \citep[see details in, e.g.,][]{ballester2018}. Alfvén waves are incompressible transverse waves driven by the magnetic tension force. We neglected all non-ideal terms in the basic single-fluid MHD equations, with the exception of Ohm's and ambipolar diffusion. In order to study linear Alfvén waves, we linearize the equations by writing each variable as the sum of the background value and a small perturbation. Then, we only retain linear terms in the perturbations. After some calculations, the linearized equations that govern Alfv\'en waves are the transverse components of the momentum and induction equations, namely:
\begin{eqnarray}
    \rho\frac{\partial v_{\perp}}{\partial t} &=& \frac{B}{\mu_0}\frac{\partial B_{\perp}}{\partial z}, \label{eq:moment} \\
     \frac{\partial B_{\perp}}{\partial t} &=& B \frac{\partial v_{\perp}}{\partial z}+  \frac{\partial}{\partial z} \left( \eta_{\rm C} \frac{\partial B_{\perp}}{\partial z}\right), \label{eq:induct}
\end{eqnarray}
where $v_{\perp}$ and $B_{\perp}$ are the velocity and magnetic field perturbations perpendicular to the background magnetic field, respectively, and $\eta_{\rm C} = \eta + B^2 \eta_{\rm A}$ is Cowling's (or total) diffusion coefficient. We plot in Figure~\ref{fig:magnitudes}f Cowling's coefficient, which varies many orders of magnitude along the thread. In the cool partially ionized region, $\eta_{\rm C} $ is dominated by the ambipolar coefficient  multiplied by the magnetic field strength squared, whereas in the hot fully ionized region $\eta_{\rm C} = \eta$.

We study Equations~(\ref{eq:moment}) and (\ref{eq:induct}) in considering the stationary state of wave propagation. In this situation, it is assumed that the driver that excites the waves has been working for a sufficiently long time that is, at least, twice the Alfv\'en travel time along the thread. So, we can express the temporal dependence of the perturbations as $\exp \left( - i \omega t  \right)$, where $\omega$ is the angular wave frequency. The linear wave frequency is simply $f=\omega/2\pi$. Hence, we can combine Equations~(\ref{eq:moment}) and (\ref{eq:induct}) to obtain a single partial differential equation for  $B_{\perp}$ alone, namely
\begin{equation}
 \frac{\partial^{2} B_{\perp}}{\partial z^{2}} 
        + \frac{\frac{\partial}{\partial z} \left(\va^2 - i \omega \eta_{\rm C}  \right)}{\left(\va^2 - i \omega \eta_{\rm C}  \right)}
        \frac{\partial B_{\perp}}{\partial z} + \frac{\omega^2}{\left(\va^2 - i \omega \eta_{\rm C}  \right)} B_{\perp}=0, \label{eq:main}
\end{equation}
where $\va = B/\sqrt{\mu_0 \rho}$ is the Alfv\'en speed. In turn, $v_{\perp}$ is related to $B_{\perp}$ as
\begin{equation}
    v_{\perp} = \frac{i}{\omega} \frac{\va^2}{B} \frac{\partial B_{\perp}}{\partial z}. \label{eq:velocity}
\end{equation}
Equation~(\ref{eq:main}) is our main equation and will be integrated using standard numerical methods after assuming boundary conditions at $z = \pm L/2$.

\subsection{Boundary conditions}
\label{sec:BC}

To numerically solve Equation~(\ref{eq:main}), we assume that waves are driven at $z=-L/2,$ with a prescribed amplitude and a given frequency, $f$. The driver imposes a magnetic field perturbation amplitude of the form:
\begin{equation}
B_\perp = A \left(f\right) \exp\left( i \Phi \left(f\right) \right) \qquad \textrm{at} \qquad z = -\frac{L}{2},
\end{equation}
where $A \left(f\right)$ is the spectral weighting function and $0 < \Phi \left(f\right) < 2\pi$ is a random phase different for each frequency. The result of waves excited by a broadband driver is constructed by varying the driver frequency in a wide range and superposing the solutions obtained for individual, discrete frequencies. We consider 1001 discrete driver frequencies between 0.1 and 1,000~mHz in logarithmic spacing. The superposition is done after assuming a particular form for $A \left(f\right)$. The  spectral weighting function for Alfv\'en waves in solar prominences is largely unknown. Apart from  \citet{hillier2013}, who performed a statistical study of transverse oscillations in a quiescent prominence, no other observational work has addressed this issue. Here, the spectral weighting function is assumed to be of a double power-law form, namely:
\begin{equation}
A\left(f\right) = A_0 \left\{ \begin{array}{lll}
\left(\frac{f}{f_b}\right)^{1/4}, & \textrm{if} & f \leq f_b, \\
\left(\frac{f}{f_b}\right)^{-5/6}, & \textrm{if} & f > f_b,
\end{array} \right. \label{eq:spectral}
\end{equation}
where $f_b$ is the break frequency of the spectrum and $A_0$ is a constant. The low-frequency exponent of $1/4$ is considered after the results of \citet{hillier2013}. They showed that for periods, $P$, larger than 50~s, the velocity amplitude of the  transverse oscillations approximately scales as $\sim P^{-0.25 \pm 0.04}$. We set the break frequency to $f_b = 20$~mHz, which corresponds to the smallest period reported by \citet{hillier2013} owing to instrumental constrains. Conversely, since we have no information about the scaling for higher frequencies, we simply assume a Kolmogorov-like exponent of $-5/6$. Finally, the value of the constant $A_0$ depends on the prescribed value of the driven energy flux. 

For an Alfvén wave, the energy flux averaged over one full period is given by \citep[see][]{walker2004}:
\begin{equation}
    \langle \boldsymbol{\pi} \rangle = -\frac{1}{2\mu_{0}} \textrm{Re} \left[ v_{\perp} B_{\perp}^{*} \right] \mathbf{B},
\end{equation}
where * denotes the complex conjugate. The energy flux can be separated into the parallel and anti-parallel contributions with respect to the magnetic field direction, which are denoted by $\langle \boldsymbol{\pi} \rangle^{\uparrow}$ and $\langle \boldsymbol{\pi} \rangle^{\downarrow}$, respectively. In order to write these contributions, we use the Elsässer variables \citep{elsasser1950}, which are defined as
\begin{eqnarray}
    Z^{\uparrow} = v_{\perp} - \frac{B_{\perp}}{\sqrt{\mu_{0}\rho}}, \label{eq:elsup} \\
    Z^{\downarrow} = v_{\perp} + \frac{B_{\perp}}{\sqrt{\mu_{0}\rho}}, \label{eq:elsdown}
\end{eqnarray}
where $Z^{\uparrow}$ and $Z^{\downarrow}$ represent parallel-propagating and anti-parallel-propagating Alfv\'enic disturbances, respectively. With these new variables, we can write the energy flux as $\langle \boldsymbol{\pi} \rangle = \langle \boldsymbol{\pi} \rangle^{\uparrow} - \langle \boldsymbol{\pi} \rangle^{\downarrow}$, where
\begin{eqnarray}
    \langle \boldsymbol{\pi} \rangle^{\uparrow} &=& \frac{1}{8} \sqrt{\frac{\rho}{\mu_{0}}} Z^{\uparrow}  Z^{\uparrow *}  \vec{B}, \\
    \langle \vec{\pi} \rangle^{\downarrow} &=& \frac{1}{8} \sqrt{\frac{\rho}{\mu_{0}}} Z^{\downarrow} Z^{\downarrow *}  \vec{B}.
\end{eqnarray}
The fluxes $ \langle \boldsymbol{\pi} \rangle^{\uparrow}$ and $ \langle \vec{\pi} \rangle^{\downarrow}$ represent propagation of energy in the directions parallel and anti-parallel to the magnetic field, respectively. 

At the driver location, that is, at $z=-L/2$, $ \langle \boldsymbol{\pi} \rangle^{\uparrow}$ corresponds to the energy flux injected by the driver, while $ \langle \vec{\pi} \rangle^{\downarrow}$ is the reflected energy flux. The energy flux  of  Alfvén waves driven at the photosphere is believed to be of the order of $10^4$~W~m$^{-2}$  \citep[see, e.g.,][]{depontieu2001,goodman2011,tu2013,arber2016}. However, due to the filtering effect of the chromosphere, only about $1\%$ of the photospheric flux is capable of reaching coronal heights \citep[see a detailed study in][]{soler2019}. Therefore, we set $\sum_f  \langle \boldsymbol{\pi} \rangle^{\uparrow} = 10^2$~W~m$^{-2}$ at $z=-L/2$, which allows us to determine the value of $A_0$.

On the other hand, the boundary condition at $z=L/2$ is set according to the imposed relation between $ \langle \boldsymbol{\pi} \rangle^{\uparrow}$ and $\langle \vec{\pi} \rangle^{\downarrow} $. The paradigmatic cases are the perfectly reflecting boundary and the perfectly transparent boundary. In the perfectly reflecting boundary $\langle \vec{\pi} \rangle^{\uparrow} = \langle \vec{\pi} \rangle^{\downarrow}$, then $\langle \vec{\pi} \rangle = 0$. This condition requires that $v_{\perp} B_{\perp}^{*} =0$. If $B_{\perp}$ is arbitrary, then $v_{\perp}=0$. Using Equation~(\ref{eq:velocity}), we obtain
\begin{equation}
\frac{\partial B_{\perp}}{\partial z} = 0 \qquad \textrm{at} \qquad z = \frac{L}{2},
\end{equation}
for the perfectly reflecting boundary. In the case of a perfectly transparent boundary, we have $\langle \vec{\pi} \rangle^{\uparrow} \neq 0$ and $\langle \vec{\pi} \rangle^{\downarrow} = 0$, which is equivalent to $Z^{\downarrow}=0$. Using Equations~(\ref{eq:velocity}) and (\ref{eq:elsdown}) we obtain:
\begin{equation}
 \frac{\partial B_{\perp}}{\partial z} = \frac{i \omega}{v_{A}} B_{\perp} \qquad \textrm{at} \qquad z = \frac{L}{2},
\end{equation}
for the perfectly transparent boundary. In general, a realistic boundary condition would be neither perfectly reflective nor perfectly transparent. Our model does not include the chromosphere and photosphere where the magnetic field of prominences is actually anchored. It is the properties of the plasma in those layers that determine the reflection or transmission of the waves \citep[see, e.g.,][]{leroy1980,hollweg1981,similon1992,soler2017,soler2019}. Based on the above results,  we can define a general boundary condition as
\begin{equation}
\frac{\partial B_{\perp}}{\partial z} = \varepsilon \frac{i\omega}{v_{A}} B_{\perp} \qquad \textrm{at} \qquad z = \frac{L}{2},
\label{eq:param}
\end{equation}
where $\varepsilon \in \left[0,1 \right]$ is a parameter. Checking the extreme values of $\varepsilon$ we recover the perfectly reflecting boundary for $\varepsilon=0$ and the perfectly transparent boundary for $\varepsilon=1$. Values of $\varepsilon$ in between those limit cases correspond to intermediate situations.

\subsection{Wave absorption and heating rate}

We define the reflectivity, $\mathcal{R}$, and transmissivity, $\mathcal{T}$, coefficients, which physically describe  the fraction of the driven energy that is reflected back to the driver location or transmitted all the way through the opposite boundary of the model, respectively. We compute the incoming, reflected, and transmitted fluxes as \citep[see][]{soler2019}:
\begin{eqnarray}
    \langle \pi \rangle_{\rm inc.} &=& \left| \langle \boldsymbol{\pi}  \rangle^{\uparrow} \right| \qquad \textrm{at} \qquad z = -\frac{L}{2}, \\
    \langle \pi \rangle_{\rm ref.} &=& \left| \langle \boldsymbol{\pi}  \rangle^{\downarrow} \right| \qquad \textrm{at} \qquad z = -\frac{L}{2}, \\
    \langle \pi \rangle_{\rm tra.} &=& \left| \langle \boldsymbol{\pi}  \rangle \right| \qquad \textrm{at} \qquad z = \frac{L}{2}.
\end{eqnarray}
With the help of these fluxes, the coefficients are readily computed as
\begin{eqnarray}
    \mathcal{R} &=& -\frac{\langle \pi \rangle_{\rm ref.}}{\langle \pi \rangle_{\rm inc.}},\label{eq:r} \\
    \mathcal{T} &=& \frac{\langle \pi \rangle_{\rm tra.}}{\langle \pi \rangle_{\rm inc.}}. \label{eq:t}
\end{eqnarray}
In a  stationary system with no dissipation, $\mathcal{R} + \mathcal{T} = 1$ because of energy conservation. However, the presence of dissipation causes $\mathcal{R} + \mathcal{T} < 1$. Then, we can define the absorption coefficient, $ \mathcal{A}$, which represents  the fraction of the driven wave energy that is absorbed or deposited in the plasma because of dissipation, namely:
\begin{equation}
    \mathcal{A} = 1 - \left( \mathcal{R} + \mathcal{T} \right). \label{eq:a}
\end{equation}
These three coefficients are functions of the driver frequency but are independent of the driver amplitude. They are intrinsic properties of the model.

\begin{figure*}[!tb]
    \centering
    \includegraphics[width=9cm,height=6cm]{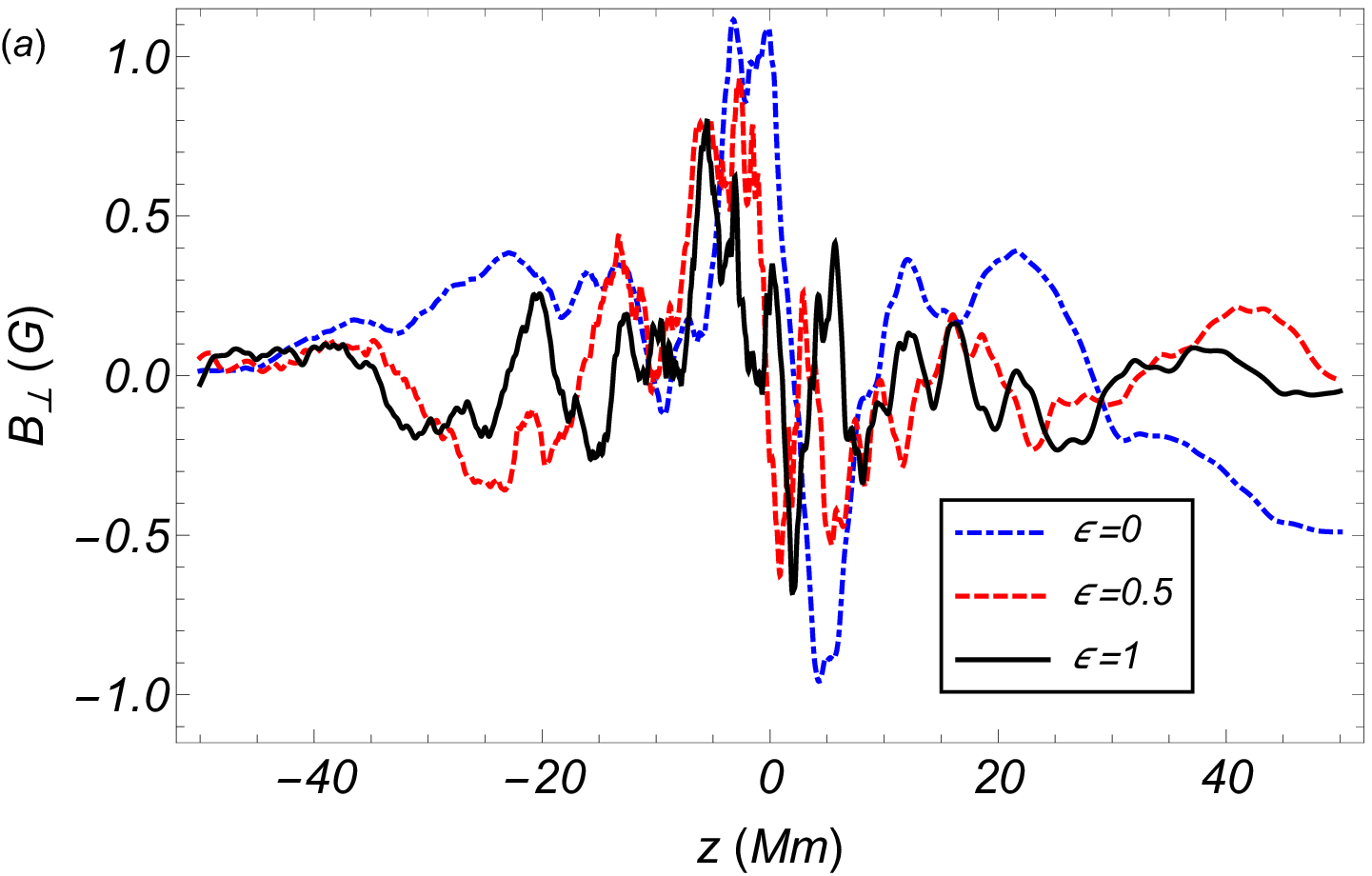} 
    \includegraphics[width=9cm,height=6cm]{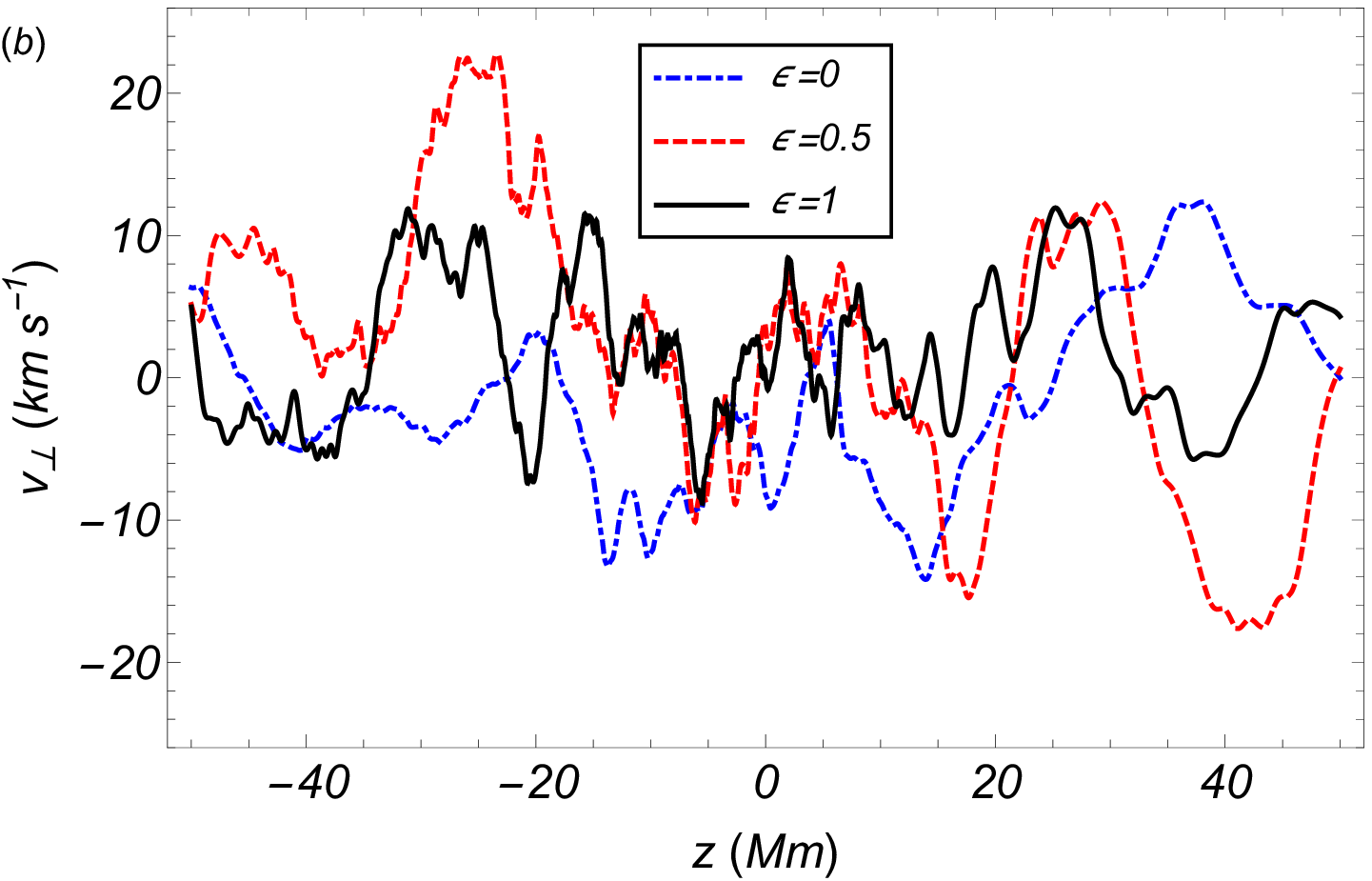} 
    \includegraphics[width=9cm,height=6cm]{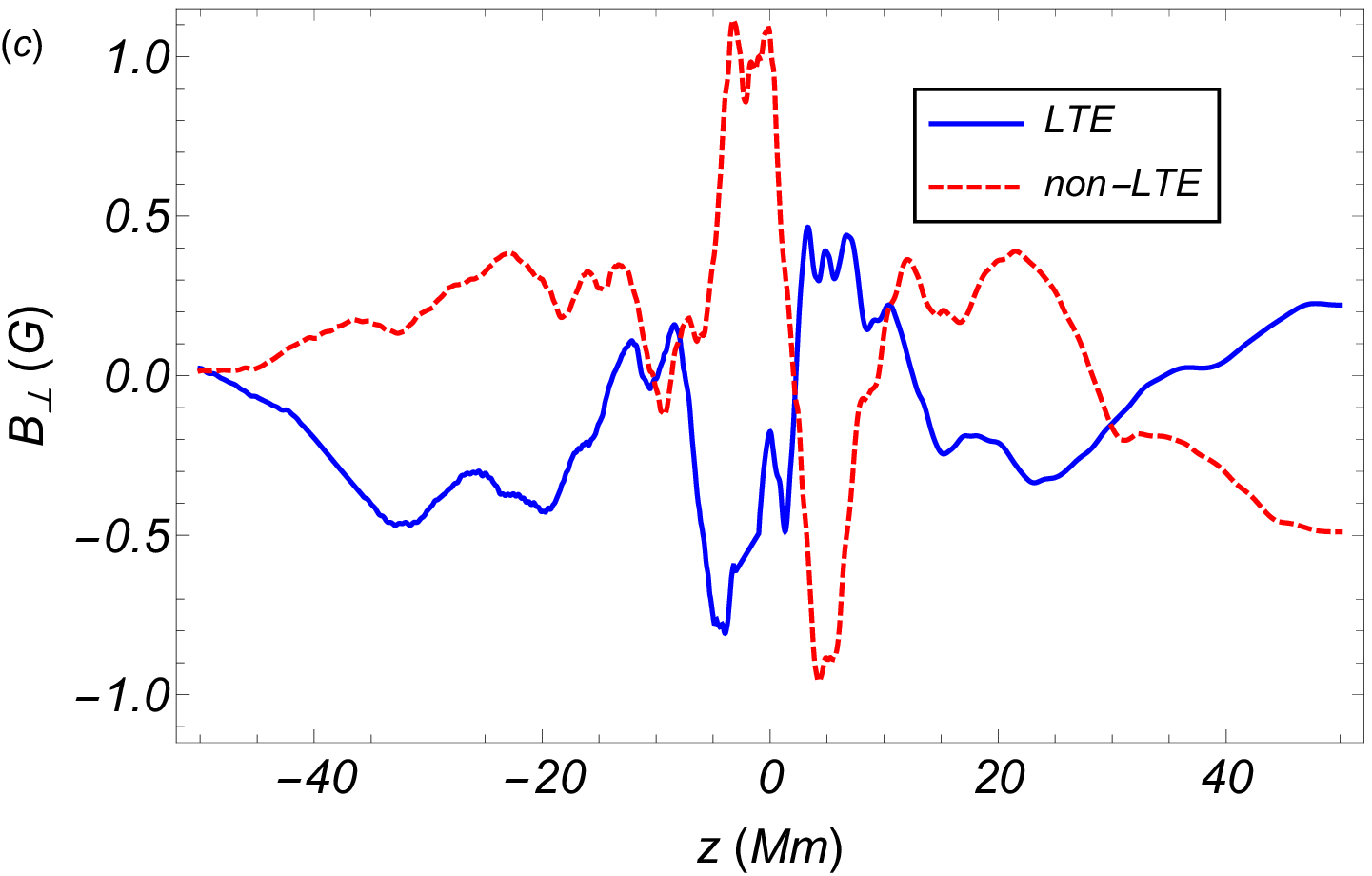} 
    \includegraphics[width=9cm,height=6cm]{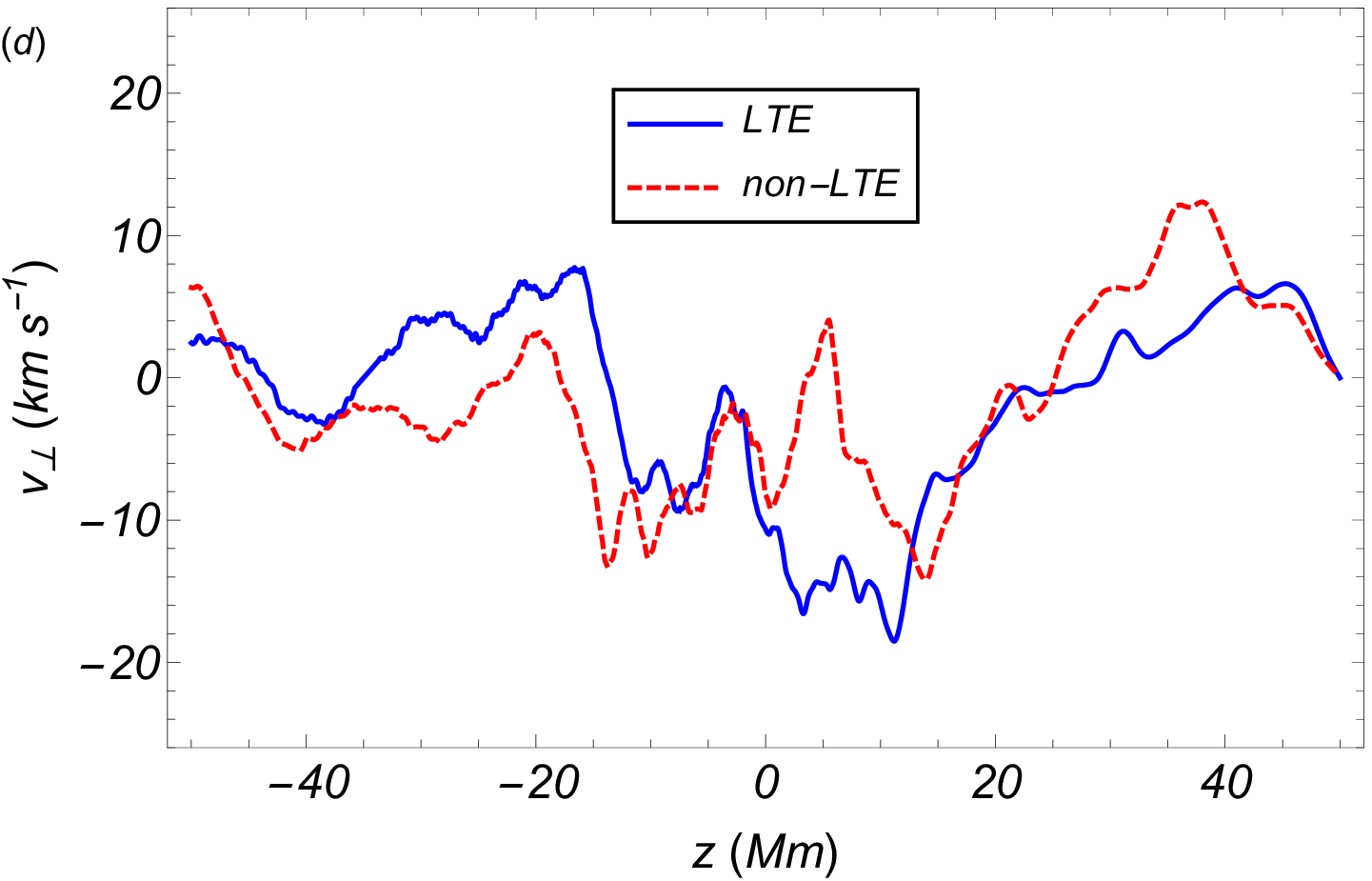} 
    \caption{Real part of the \textbf{a)} magnetic field and \textbf{b)} velocity perturbations along the thread for the non-LTE case with $\varepsilon=$~0, 0.5, and 1; \textbf{c)} and \textbf{d)} are same as \textbf{a)} and \textbf{b)} but with $\varepsilon=0$ and comparing the results for LTE and non-LTE cases.}
    \label{fig:field}
\end{figure*}

The consequence of the wave energy absorption is the heating of the plasma. The heating is calculated using the Joule heating function, which represents the dissipation of electric currents  associated with the Alfv\'en waves. The  heating function, $Q$, can be written as
\begin{equation}
   Q = \mu_{0}\eta \left| \vec{j_{\parallel}} \right|^{2} + \mu_{0} \eta_{\rm C} \left| \vec{j_{\perp}} \right|^{2},
\end{equation}
where $\vec{j}_\parallel$ and $\vec{j}_\perp$ are the perturbations of the parallel and perpendicular components of the electric current with respect to the background magnetic field. We see that Cowling's diffusion, that is, the joint effect of Ohm's and ambipolar diffusion, is responsible for the dissipation of perpendicular currents, whereas   parallel currents are dissipated by Ohm's diffusion alone \citep[see, e.g.,][]{khomenko2012}. In our case, we have that
\begin{eqnarray}
\vec{j}_\parallel &=& 0, \\
\vec{j}_\perp &=&  \frac{1}{\mu_0} \frac{\partial B_{\perp}}{\partial z} \hat{e}_\perp,
\end{eqnarray}
where $\hat{e}_\perp$ denotes the unit vector in the perpendicular direction. There are no parallel currents in our model, so that all the heating is produced by the dissipation of perpendicular currents. Therefore, the heating function averaged over one full period of the wave is:
\begin{equation}
     \langle Q \rangle = \frac{\eta_{\rm C}}{2\mu_{0}}\left|\frac{\partial B_{\perp}}{\partial z} \right|^{2}. \label{eq:heating}
\end{equation}
 The computation of the heating rate in the case of a broadband spectrum is done in the same fashion as for the energy flux. Namely, the heating function is calculated for all the individual frequencies in the spectrum using Equation~(\ref{eq:heating}) and then the results are added together according to the spectral weighting function.

\section{Results}
\label{sec:results} 
 
\subsection{Magnetic field and velocity perturbations}
\label{sec:perts}

We start by studying the magnetic field and velocity perturbations along the thread because of the presence of the broadband spectrum of Alfv\'en waves. These results are displayed in Figure~\ref{fig:field}. The perturbations are complex quantities and  we plot their real part only. 

Figures~\ref{fig:field}a and \ref{fig:field}b compare the perturbations for $\varepsilon=$~0, 0.5, and 1 in the non-LTE case.  Before discussing these results, we note that the random phase assumed during superposition of the different frequencies in the spectrum is different for the  three values of $\varepsilon$. This implies that the fine details of the obtained perturbations are also intrinsically different.

The driven broandband spectrum naturally results in a mixture of many spatial scales (wavelengths) along the thread. However, it is evident from Figures~\ref{fig:field}a and \ref{fig:field}b that smaller wavelengths are present in the central, denser part of the thread than in the surrounding evacuated zones. The reason for this result is that the density enhancement at the thread center causes the local decrease of the Alfv\'en speed. The values of the Alfv\'en  speed at the tube center and at its ends are $\sim$~\SI{89}{\km\per\second} and $\sim$~\SI{892}{\km\per\second}, respectively. For an Alfv\'en wave, the wavelength is proportional to the  Alfv\'en  speed. The consequence of this is the decrease of the wavelength in the central part \citep[for a similar result but in the case of chromospheric waves see, e.g.,][]{depontieu2001,zaqarashvili2013,soler2017}.

The amplitude of the magnetic field perturbation is larger at the thread center than at the ends, while the opposite behaviour is found in the case of the velocity perturbation.  Again, this can be explained by the dependence of the density along the thread and the fact that in an Alfv\'en wave there is equipartition between kinetic and magnetic energies \citep{ferraro1958}. The maximum value of the magnetic field perturbation found near the thread center is  $\sim$~\SI{1}{\gauss}, which is one order of magnitude below the equilibrium magnetic field strength. In turn, the maximum values of the velocity found near the tube ends are  $\sim$~\SI{25}{\km\per\second}, which is again much smaller than the local Alfv\'en  speed. Therefore, the obtained amplitudes justify the use of linearized theory. In addition, the obtained velocity amplitudes are of the same order as the observationally reported Doppler velocities in small-amplitude prominence oscillations \citep[see the review by][]{arregui2018}, which indicates that the considered influx of energy by the driver is consistent with the observed amplitudes.

The results obtained for the three different values of $\varepsilon$ are qualitatively similar. The case with $\varepsilon = 0$, that is, total reflection at the right boundary, is the one that differs most from the others. For this case, larger magnetic field perturbations are found at the right end. Clearly, the condition that the waves are forced to reflect at the right boundary has an impact on the magnetic field perturbation. Also, slightly larger velocity perturbations are obtained in the evaluated part of the tube when $\varepsilon = 0$. There are no other outstanding differences.

On the other hand, Figures \ref{fig:field}c and \ref{fig:field}d compare the perturbations obtained for the LTE and non-LTE cases for the particular value of $\varepsilon = 0$.  Both magnetic field and velocity perturbations have similar behaviors in the two cases, as already discussed above. The main difference between the two cases resides in the velocity perturbation, which is larger for the non-LTE case near the tube ends. It is difficult to determine whether this disparity is actually related to the different profiles used in the LTE and non-LTE cases or, on the contrary, it is simply a consequence of the different random phases considered in the superposition of  frequencies.

\begin{figure*}[!htb]
    \centering
    \includegraphics[width=6cm,height=4cm]{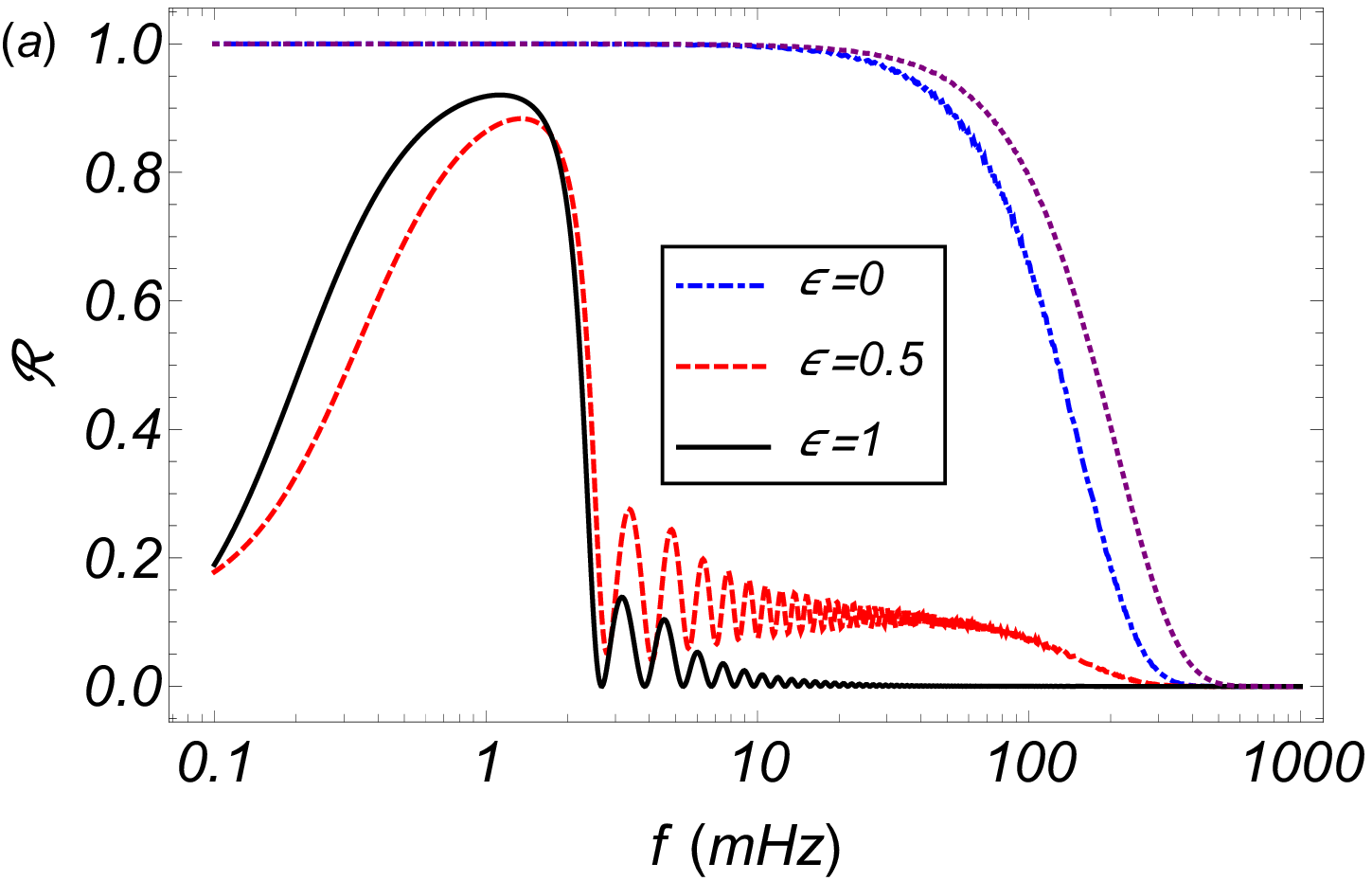} 
    \includegraphics[width=6cm,height=4cm]{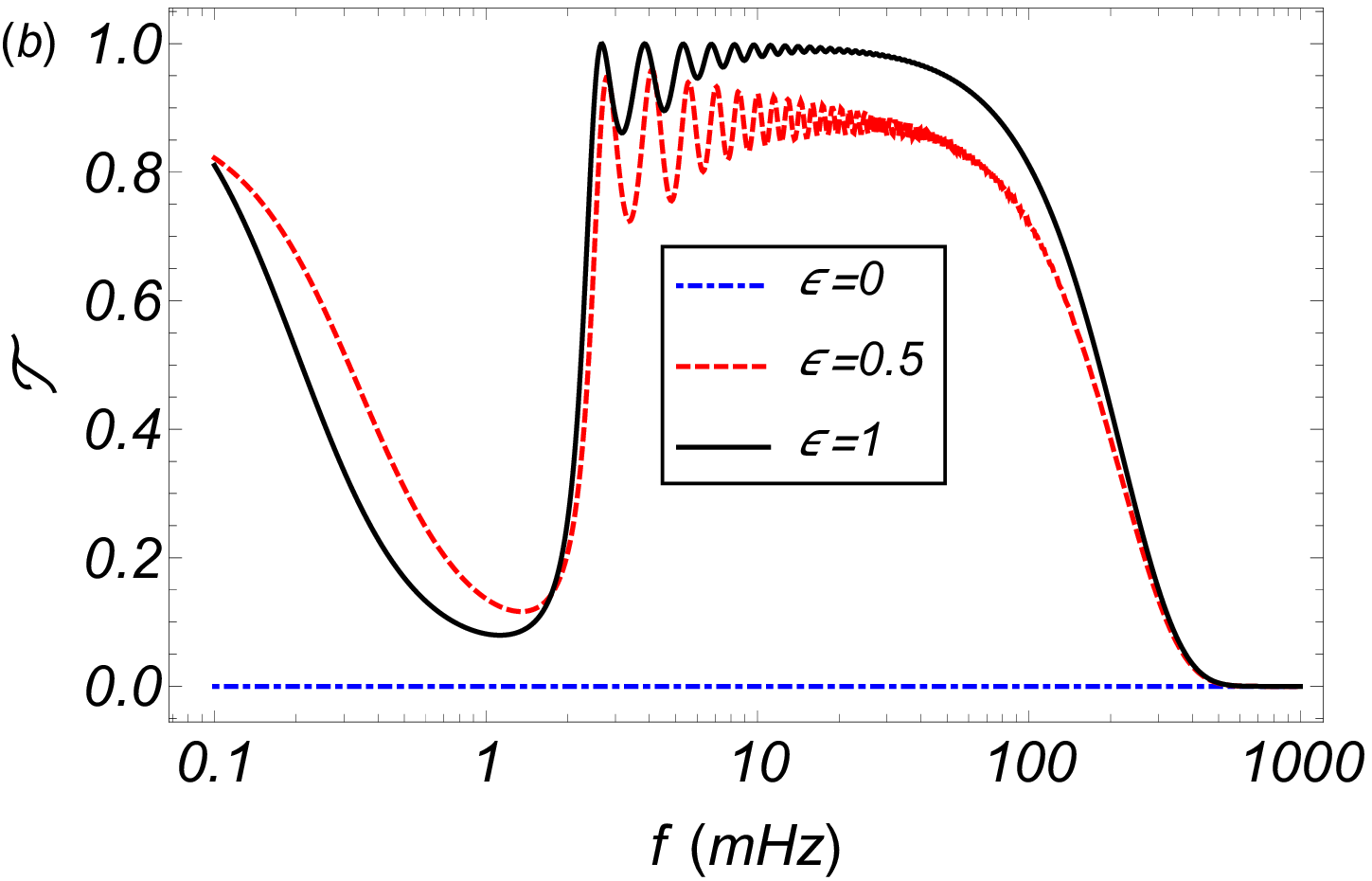} 
    \includegraphics[width=6cm,height=4cm]{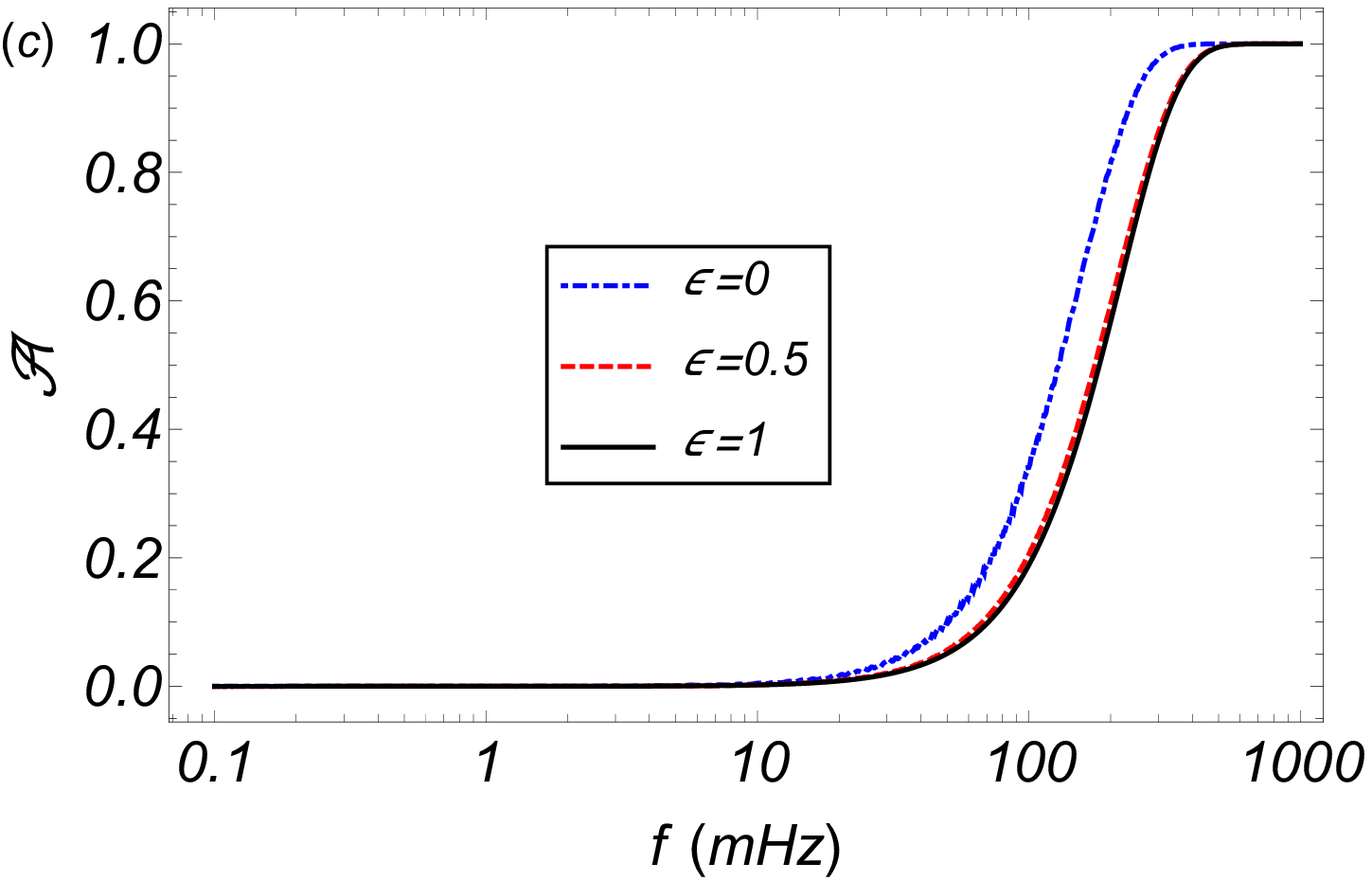} 
    \caption{Reflectivity \textbf{(a);}  transmissivity \textbf{(b)}; and absorption as functions of the frequency \textbf{(c)} in log scale for $\varepsilon=$~0, 0.5, and 1 in the non-LTE case. The purple dotted line in \textbf{(a)} corresponds to the sum of reflectivity and transmissivity for $\varepsilon=$~0.5.}
    \label{fig:absh}
\end{figure*}

\begin{figure}[htbp]
    \resizebox{\hsize}{!}{\includegraphics{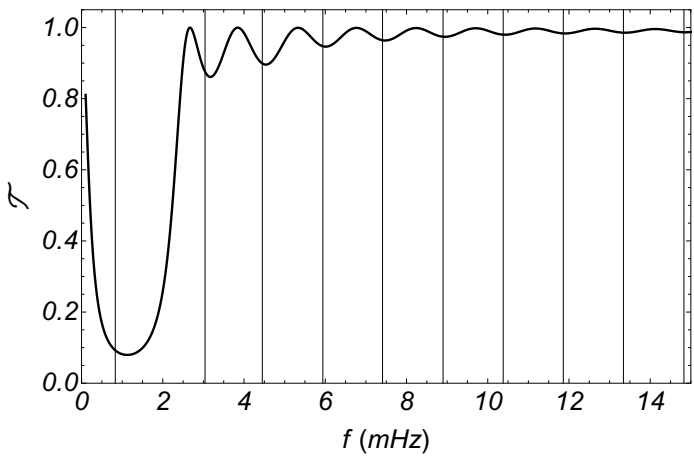}}
    \caption{Transmissivity coefficient in the non-LTE case for $\varepsilon=1$ in the range of frequencies corresponding to the first ten standing modes of the thread (vertical black lines).}
    \label{fig:res}
\end{figure}

\begin{figure*}[!htb]
    \centering
    \includegraphics[width=6cm,height=4cm]{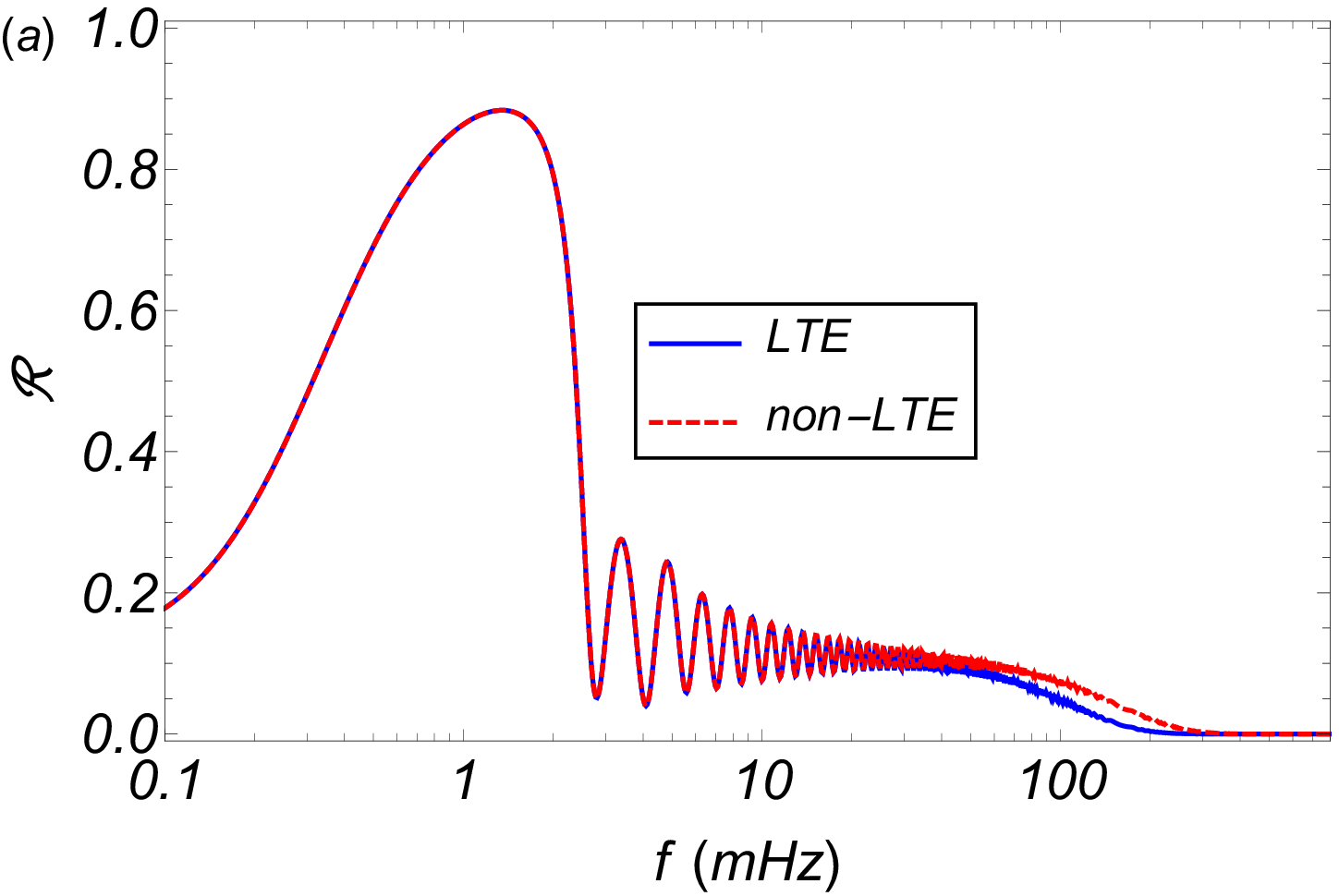} 
    \includegraphics[width=6cm,height=4cm]{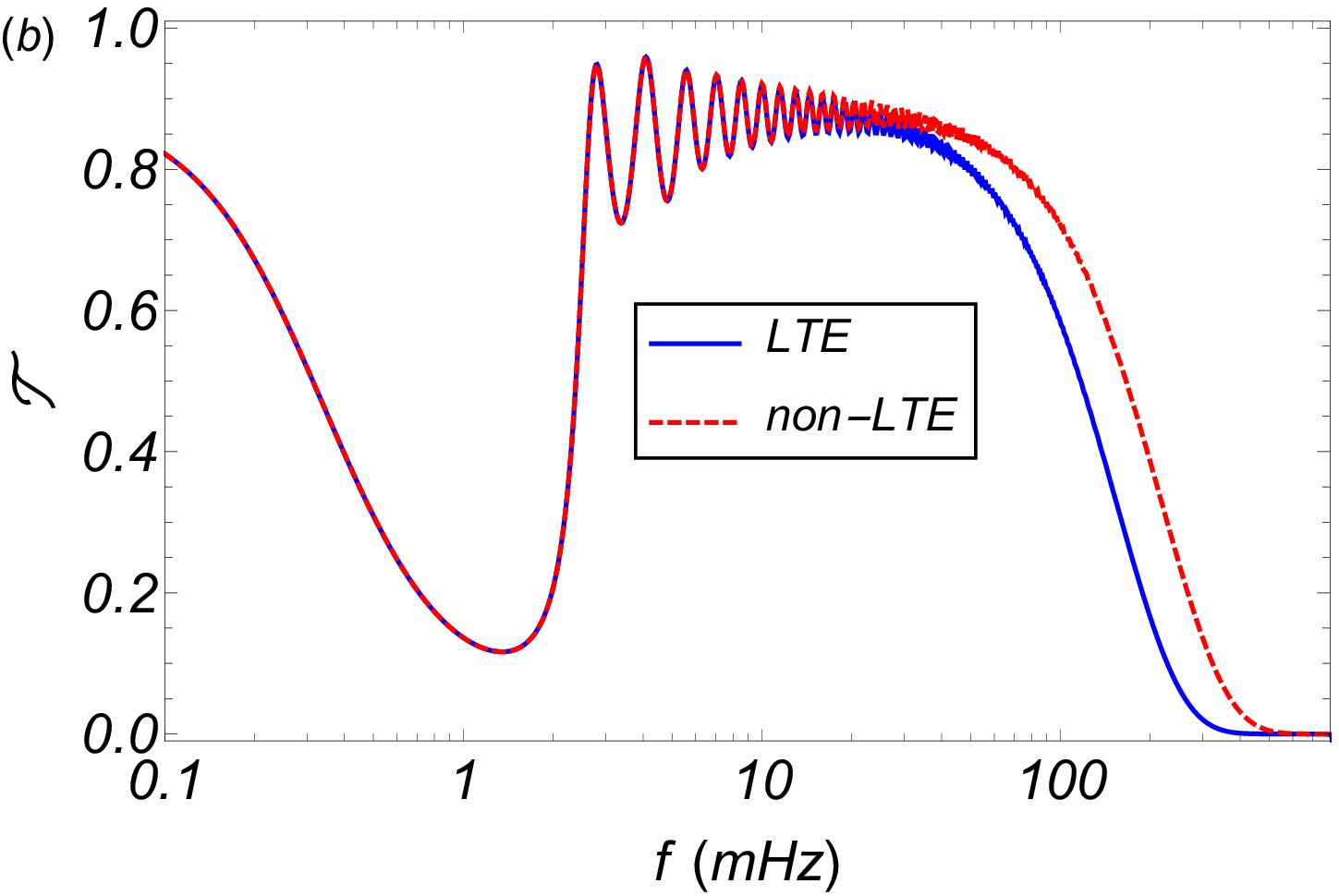} 
    \includegraphics[width=6cm,height=4cm]{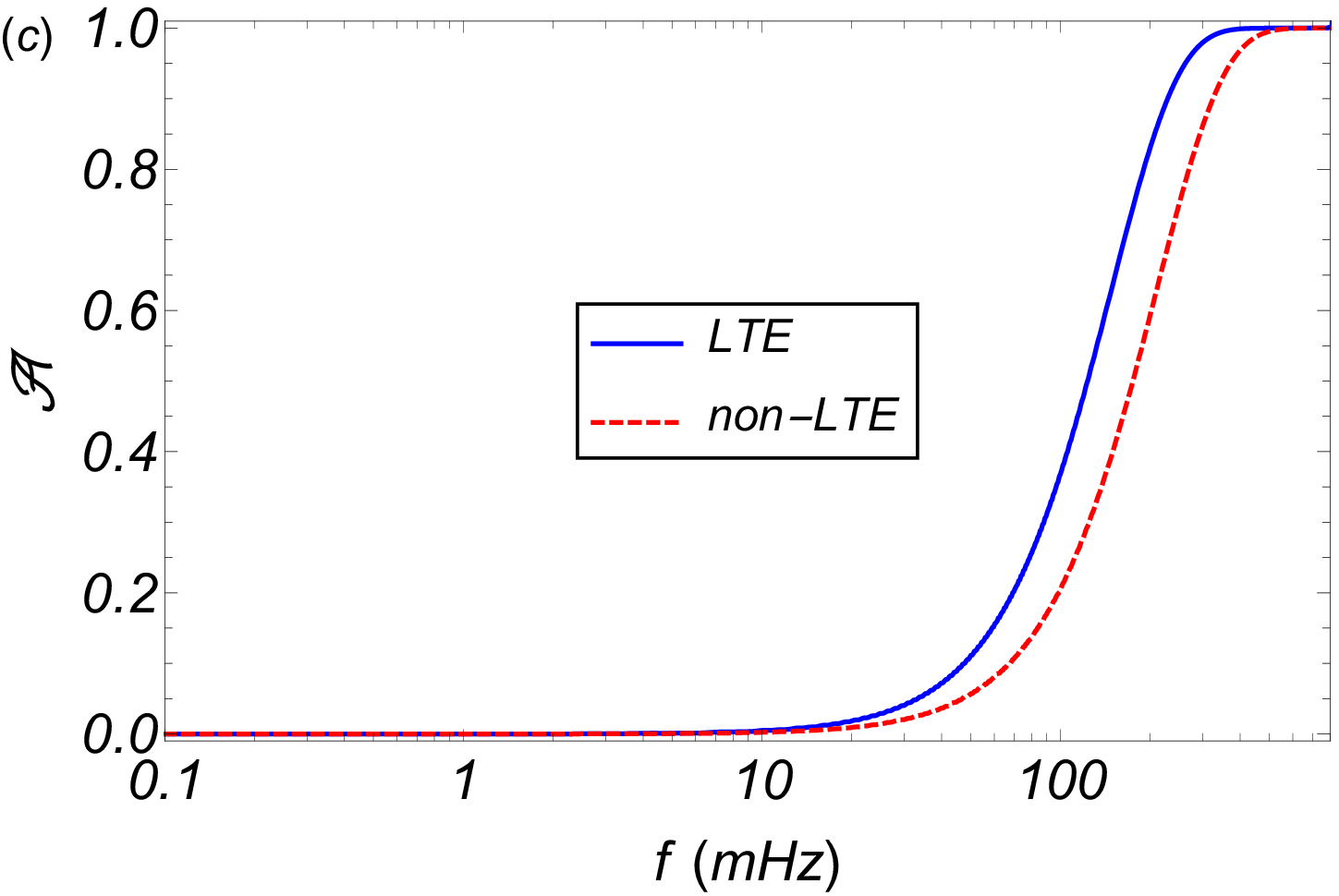} 
    \caption{Same as Figure \ref{fig:absh} but for a comparison between  LTE and non-LTE cases with $\varepsilon=0.5$.}
    \label{fig:abss}
\end{figure*}

\subsection{Reflectivity, transmissivity, and absorption}
\label{sec:coefs}

Here, we show the computations of the coefficients of reflectivity, $\mathcal{R}$, transmissivity, $\mathcal{T}$, and absorption, $\mathcal{A}$. These coefficients are calculated using Equations~(\ref{eq:r})--(\ref{eq:a}).

In Figure \ref{fig:absh} we plot the three coefficients as functions of the driver frequency, $f$,  for the non-LTE case and compare the results of  $\varepsilon=$~0, 0.5, and 1. The reflection and transmission coefficients, shown in Figures~\ref{fig:absh}a and \ref{fig:absh}b respectively,   are remarkably different for $\varepsilon= 0$ from those obtained for $\varepsilon=$~0.5 and 1, so we start by discussing the results with $\varepsilon= 0$. In this case, $\mathcal{T} = 0$ because of the imposed boundary condition: the waves are not allowed to be transmitted through the right end of the tube. As a consequence, the reflectivity is found to be maximal, that is, $\mathcal{R} \approx 1$, for $f \lesssim 10$~mHz, approximately. There is no energy deposition into the plasma for those low frequencies since all the energy returns back to the driver location. For higher frequencies the reflectivity starts to decrease as a result of the increase of the absorption (see Figure~\ref{fig:absh}c). For $f \gtrsim 300$~mHz, approximately, $\mathcal{A} \approx 1$, so that all the wave energy is absorbed into the thread plasma. These findings are consistent with the well-known result that the efficiency of the damping due to Cowling's diffusion increases with the wave frequency \citep[see the review by][and references therein]{ballester2018}.

The behavior of the reflection and transmission coefficients when $\varepsilon=$~0.5 and 1 is very different from that when $\varepsilon= 0$. The most obvious disparity is the nonzero value of $\mathcal{T}$. Indeed, the transmission coefficient is found to be significantly larger than the reflection coefficient in most parts of the frequency range, which indicates that the condition of no transmission imposed when $\varepsilon= 0$ affects dramatically wave behavior. As expected,  the larger $\varepsilon$, the larger $\mathcal{T}$ for a given frequency.  Another remarkable result is the nonmonotonic behavior of  $\mathcal{R}$ and $\mathcal{T}$ for low and intermediate frequencies. Both coefficients display oscillations, with the first one being of very large amplitude. In the appendix, we show that the presence of these oscillations is caused by resonances associated with the standing modes of the thread. The oscillations in $\mathcal{T}$ are highlighted in Figure~\ref{fig:res} for a particular case. The oscillations of the coefficient have their minimums at the standing mode frequencies, approximately, which are indicated by vertical lines. This clearly confirms that the nature of the oscillations is related to the existence of resonances  with the standing modes. Physically, the result that $\mathcal{T}$ is minimum and  $\mathcal{R}$ is maximum at the standing mode frequencies can be explained by the fact that a standing wave, represented as the superposition of two waves propagating in opposite directions, has a null net energy flux. We refer readers to the appendix for a study of the standing modes.

 Putting the oscillations aside, the global trend of the coefficients shows the decrease (increase) of the reflectivity (transmissivity) with the frequency, until the absorption starts to play a role for sufficiently high frequencies. No oscillations are present once the absorption begins to rise. This is so because dissipation starts to dominate and diminishes the role of the resonances \citep[see also][]{soler2016}. Thus, regarding the oscillations in the reflectivity or transmissivity, they are not present for frequencies lower than those of the standing mode frequencies because of the absence of resonances. They are neither present for very high frequencies because dissipation is strong enough to suppress the oscillations. 
 
 Concerning the absoption coefficient, there are no important differences between the results for various values of $\varepsilon$.  The case with $\varepsilon= 0$ reaches  total absorption, that is, $\mathcal{A} \approx 1$, faster than the other cases, which is consistent with the fact that, in this situation, no fraction of wave energy can escape through the right end of the thread, and so there is more energy available in the system to be dissipated. 

To further understand why for $\varepsilon=0$ there are no oscillations related to resonances with standing modes in the reflectively, we have overplotted in Figure \ref{fig:absh}a the sum of reflectivity and transmissivity, namely $\mathcal{R} + \mathcal{T}$, for the $\varepsilon= 0.5$ case. Surprisingly, $\mathcal{R} + \mathcal{T}$ does not display oscillations and the result matches quite well with the reflectively of the $\varepsilon=0$ case. The reason is that when $\varepsilon \neq 0$ the oscillations in $\mathcal{R}$ are out of phase with respect to those of $ \mathcal{T}$ in a fashion so that they cancel out when $\mathcal{R}$ and  $\mathcal{T}$ are added. This implies that resonances must surely be present in the $\varepsilon =0$ case too, but the reflectively is not the appropriate parameter to show them. As shown later in Section~\ref{sec:heating}, resonances clearly appear in the integrated heating rate when $\varepsilon=0$.

\begin{figure*}[!htb]
    \centering
    \includegraphics[width=9cm,height=6cm]{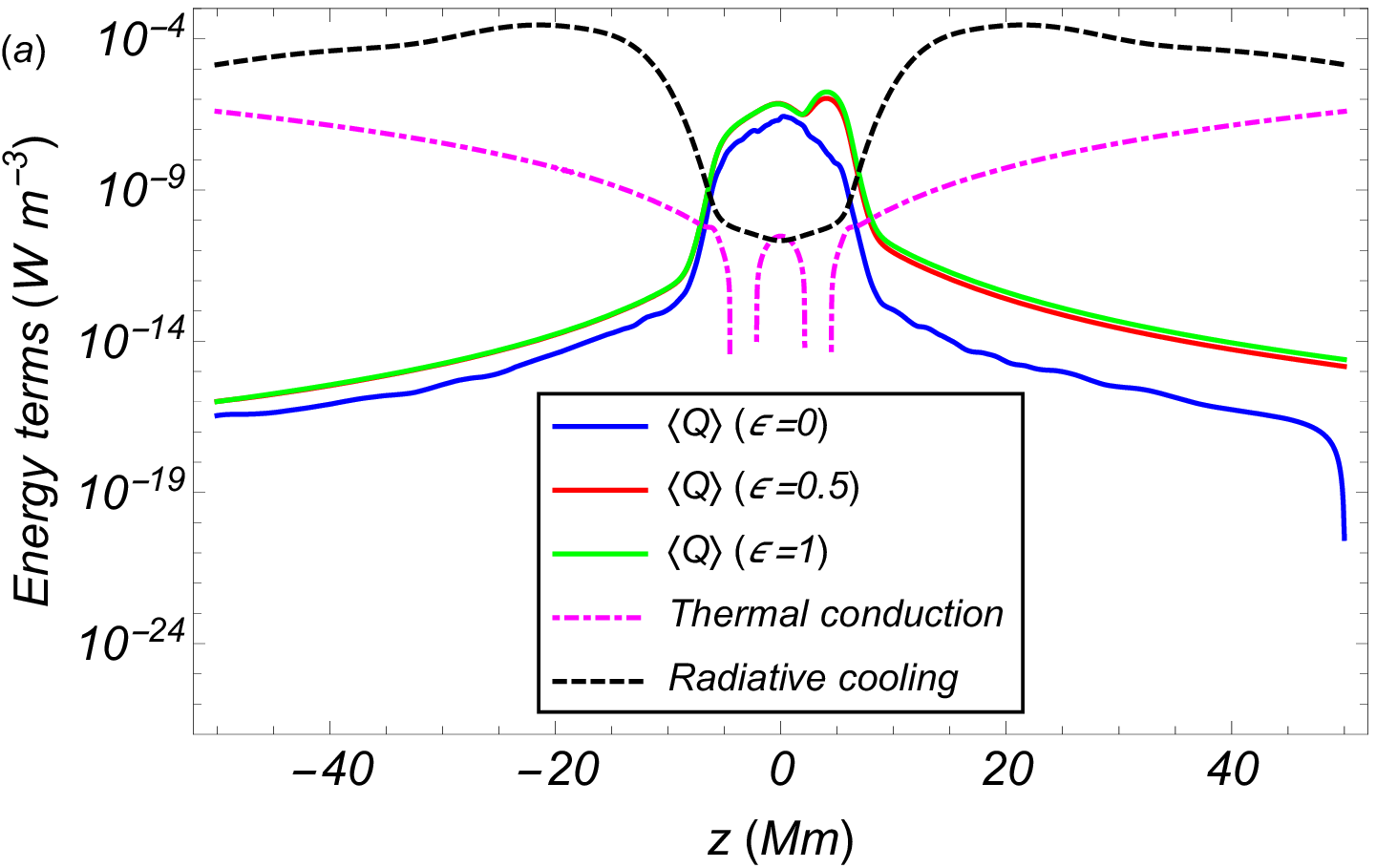} 
    \includegraphics[width=9cm,height=6cm]{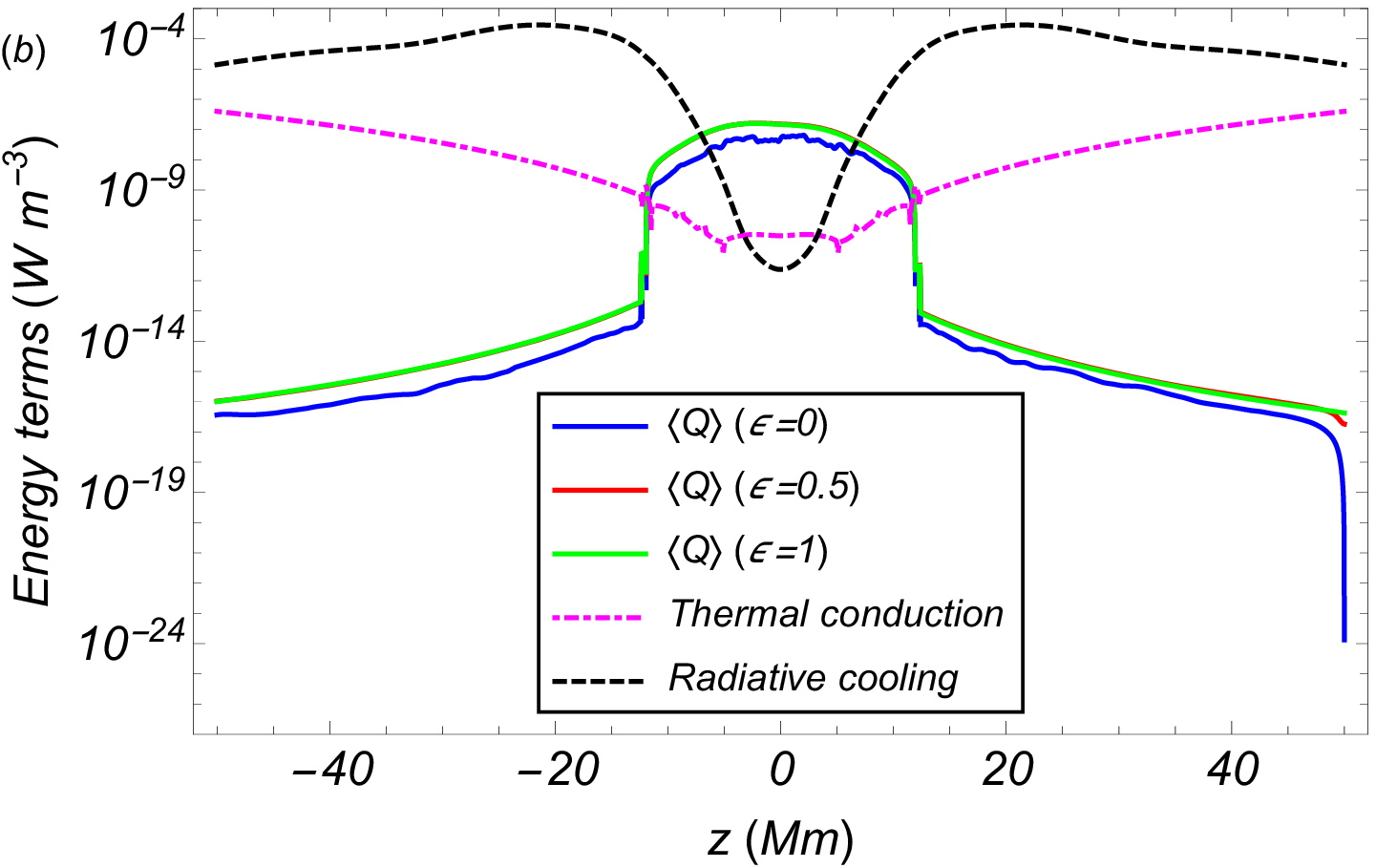} 
    \caption{Wave heating function (for $\varepsilon=$~0, 0.5, and 1), cooling function, and thermal conduction term  along the thread for the \textbf{a)}  LTE and  \textbf{b)} non-LTE cases.}
    \label{fig:balance}
\end{figure*}

The results discussed above correspond to the non-LTE case. In Figure~\ref{fig:abss}, we compare the three coefficients obtained in the LTE and non-LTE cases for $\varepsilon = 0.5$. The reflectivity is almost identical for both cases in the whole frequency range. The transmissivity is also almost the same for low and intermediate frequencies until the absorption begins the grow. In this model, the wave reflection and transmission is essentially governed by the longitudinal variation of density, which is the same in both cases. This explains why the reflectivity and transmissivity are identical in the LTE and non-LTE cases. Conversely, the wave damping depends on the values of Cowling's diffusion, which are different for the two cases. In the LTE case total absorption is reached faster than in the non-LTE case, which indicates that wave damping is more efficient in the LTE case. This is consistent with the larger values of Cowling's diffusion obtained in the LTE case (see again Figure~\ref{fig:magnitudes}f).

A remarkable difference with the results of \citet{soler2016} is that here the absorption coefficient is a smooth, monotonic function of the frequency, while in  \citet{soler2016} the absorption was characterized by the presence of narrow peaks associated with resonances with standing modes. We do find resonances in the reflection and transmission coefficient, but not in the absorption coefficient. Conversely, the shape of our absorption coefficient is very similar to that computed in the chromosphere \citep[see][]{soler2017}. In addition, the absorption values obtained by \citet{soler2016} for nonresonant frequencies were significantly smaller than the absorption computed here, so that globally the absorption of wave energy is larger in the present case. We believe that the distinct feature that causes our results to be different from those of \citet{soler2016} is the nonuniform thread model considered here, whereas   \citet{soler2016} considered a uniform prominence slab embedded in a uniform corona. The piecewise model of \citet{soler2016} favors the trapping of waves within the prominence slab, and this may artificially enhance the role of resonances. A similar argument was used by \citet{soler2017} to explain the absence of resonances in their nonuniform chromospheric model as compared with the resonances obtained in  piecewise models \citep[see][]{hollweg1981}. This highlights the importance of using nonuniform models to correctly describe the reflection, transmission, and absorption properties of the waves.

\subsection{Heating rates}
\label{sec:heating}

The total volumetric heating rates associated with wave dissipation are plotted in Figure~\ref{fig:balance}. The heating profiles for the LTE and non-LTE profiles have almost the same global shape, which is consistent with the previous results.  Heating is maximum at the thread center and decreases by many orders of magnitude in the evacuated, coronal part. This result is related with the relative values of Cowling's diffusion coefficient in these regions. In particular, in the LTE case the central region where heating is significant is narrower than in the non-LTE case, and the maximum value of the heating rate is  slightly larger for the LTE case than for the non-LTE case. In both cases, the order of magnitude of the volumetric heating rate at the center of the thread is $\sim 10^{-7}$~W~m$^{-3}$, while at the thread ends it is $\sim 10^{-16}$~W~m$^{-3}$.

\begin{figure}[!htbp]
    \resizebox{\hsize}{!}{\includegraphics{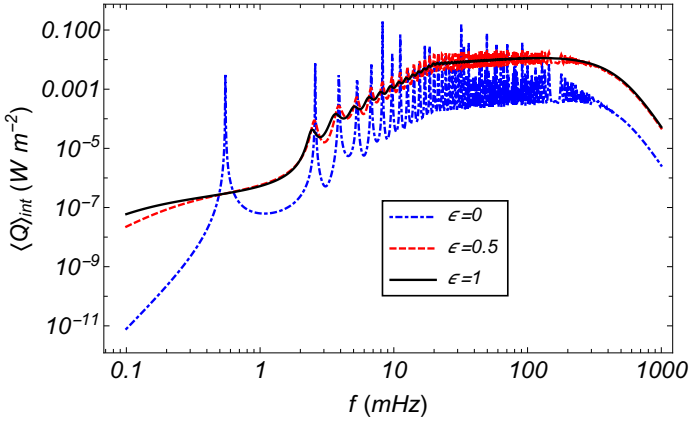}}
    \caption{Integrated heating rate along the thread as a function of the wave frequency in the non-LTE case with $\varepsilon=$~0, 0.5, and 1.}
    \label{fig:qint}
\end{figure}

The heating profiles have similar forms for the three different boundary conditions considered. In the cases of total and partial transmission, the results almost overlap and are barely distinguishable in Figure~\ref{fig:balance}. Somewhat smaller heating rates are obtained in the case of total reflection ($\varepsilon = 0$). This is a counter-intuitive result, since we showed before in Section~\ref{sec:coefs} that, for the same wave frequency, the absorption coefficient is larger in the $\varepsilon = 0$ case.  In order to understand this result, we study the contribution to the heating rate of the different frequencies present in the broadband spectrum. To do so, we compute the heating rate integrated along the whole domain, namely:
\begin{equation}
    \langle Q \rangle_{\rm int} = \int_{-L/2}^{L/2} \langle Q \rangle \,{\rm d}z,
    \label{eq:int}
\end{equation}
where the integration is performed for each individual frequency in the spectrum. Figure~\ref{fig:qint} shows the  integrated heating as a function of the frequency for the three different boundary conditions in the non-LTE case. The results for the three boundary conditions display a similar trend. The heating rate increases with the frequency in the low frequency region until reaching the spectrum peak frequency at $f =$~20 mHz. Then, the heating rate saturates to a roughly constant value until it starts to decrease for $f\gtrsim 200$~mHz. This behavior can be related to the form of the spectral weighting function, which increases in the low frequency region, has its maximum at $f=$~20~mHz, and decreases afterwards. The result that for an intermediate region of frequencies, corresponding to 20~mHz~$\lesssim f \lesssim$~200~mHz, the heating rate is roughly constant despite of the decrease of the weighting function can be understood by the fact that the efficiency of  wave damping due to Cowling's diffusion increases with the frequency. These two opposite effects, namely the decrease of the spectral weighting function and the increase of the damping efficiency, seem to balance each other almost perfectly.

In addition to the global trend discussed above, the curves in Figure~\ref{fig:qint} have also some oscillations that can again be associated with resonances with the standing modes of the system (see the appendix for more). The presence of resonances in the heating rate is a result already present in the previous paper by \citet{soler2016}.  Importantly, the amplitude of these oscillations depends on the value of $\varepsilon$. When $\varepsilon=$~0.5 and 1, the oscillations have a small amplitude and they barely affect the global trend. However, in the case of total reflection ($\varepsilon = 0$) the oscillations have large amplitudes that heavily distort the shape of the curve. The fact that the presence of the resonances have a  more significant effect in the $\varepsilon = 0$ case is related to the condition that for standing modes the boundaries of the domain must behave as perfectly reflecting boundaries. The integrated heating rate in the $\varepsilon = 0$ case is  larger than that for $\varepsilon=$~0.5 and 1 between the resonant frequencies, but this only happens in very narrow intervals. Generally, the heating rate for $\varepsilon = 0$ is lower than that for $\varepsilon=$~0.5 and 1, and this happens in  wider intervals. The net result is that the total heating for $\varepsilon = 0$ is smaller, as shown in Figure~\ref{fig:balance}.

\begin{figure}[!htbp]
    \resizebox{\hsize}{!}{\includegraphics{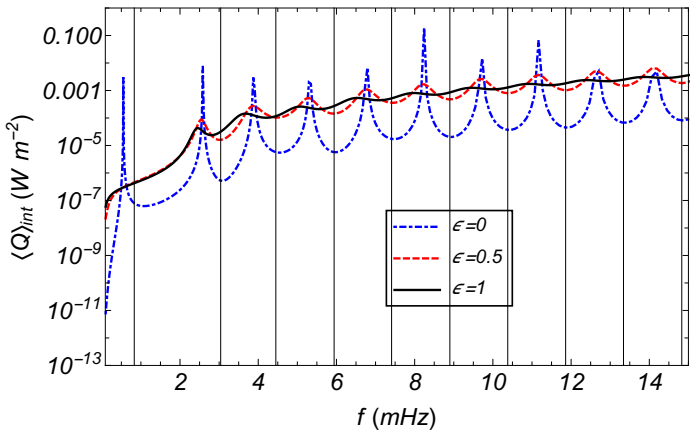}}
    \caption{Integrated heating rate along the thread as a function of the
wave frequency in the non-LTE case with $\varepsilon=$~0, 0.5, and 1. Same as Figure \ref{fig:qint} but in the range of frequencies corresponding to the first ten standing modes of the thread (vertical black lines).}
    \label{fig:qres}
\end{figure}

To shed more light on the oscillatory behavior of the integrated heating rate with the frequency, Figure~\ref{fig:qres} shows the integrated heating in the range of frequencies corresponding to the first ten standing modes. The integrated heating has relative minimums at the standing modes frequencies, which are marked with vertical lines in Figure~\ref{fig:qres}. This result reinforces the idea of the existence of resonances between the propagating waves and the standing modes of the thread, as already shown in Figure~\ref{fig:res}. The fact that the heating is minimum at the standing modes frequencies appears to stand in contradiction to the general idea of a resonance. This result is also opposite to that of \citet{soler2016}, where the maximums in the heating rate were found at the prominence slab resonance frequencies. The reason why the heating rate is minimum and not maximum at the resonance frequencies is explored in the appendix. In short, the conclusion is that the excitation of standing modes via resonances is disadvantageous with regard to the heating rate. Instead of being directly dissipated, part of the energy of the propagating waves is transferred to the standing modes at the resonance frequencies, but standing modes are very weakly damped by Cowling's diffusion and their contribution to the heating is negligible. The net result is that the heating rate is smaller than the would-be heating in the absence of resonances. In turn, the discrepancy with the results of \citet{soler2016} can be explained by the fact that in their piecewise slab model, the transmission of waves to the prominence was only efficient at the resonance frequencies because of the so-called cavity resonances \citep[see, e.g.,][]{hollweg1981}. So, in the case of \citet{soler2016} only the wave energy transmitted into the slab at the resonance frequencies was available to be dissipated efficiently. Conversely, in the present nonuniform model, wave transmission to the dense part of the thread is much more efficient than in \citet{soler2016}, as evidenced by the larger absorption obtained in the present nonuniform model. Another consequence is that the heating rates obtained here are  larger than those obtained in \citet{soler2016}.

\subsection{Discussing the energy balance}

Finally, we discuss the role that wave heating may have in the plasma energy balance. To do so, we evaluate the heat-loss function, $\mathcal{L}$, which is given by:
\begin{equation}
    \mathcal{L} = -\nabla \cdot \vec{q} - L + \langle Q \rangle + C,
    \label{eq:balance}
\end{equation}
where $\vec{q} = - \kappa \nabla T$ is the heat flux due to thermal conduction with $\kappa$ thermal conductivity, $L$ is the radiative cooling function, $\langle Q \rangle$ is the wave heating function calculated before, and $C$ is an additional unspecified source of heating. The term $C$ may represent the  heating function due to incident coronal radiation. To appropriately treat this term, we should solve the radiative transfer problem \citep[see][]{heinzel2015book}. Since this is beyond the scope of the present paper, for simplicity we set $C=0$ and omit the role of incident coronal radiation in these estimations. 

In a partially ionized plasma, the parallel thermal conductivity to the magnetic field can be approximated by $\kappa \approx \kappa_{\rm e} + \kappa_{\rm n}$, where $\kappa_{\rm e}$ and $\kappa_{\rm n}$ are the electron and neutral thermal conductivities, respectively, given by \citep[see, e.g.,][]{2015ApJ...810..146S}
\begin{eqnarray}
\kappa_{\rm e} &=& 3.2\frac{n_{e}^{2}k_{B}^{2}T}{\alpha_{e}+\alpha_{ee}}, \label{eq:neutral} \\
\kappa_{\rm n}&=&\frac{5}{3} \frac{n_{n}^{2}k_{B}^{2}T}{\alpha_{n}+\alpha_{nn}}. \label{eq:electron}
\end{eqnarray}


The cooling function is adopted after \citet{1986ApJ...308..975A}. The Athay function is frequently used in the prominence literature and may provide a more realistic representation of the cooling in the densest parts of the thread than other optically-thin cooling functions, which may overestimate radiative losses \citep[see, e.g.,][]{soler2012thermal}. Hence, we write the cooling function as:
\begin{equation}
     L(\rho,T)=f_p (T) \frac{\rho^{2} T^{2}}{m_{p}^{2}},
    \label{eq:cooling}
\end{equation}
where $m_p$ is the proton mass and $f_{p}(T)$ is an analytic function of the temperature given in  \citet{1986ApJ...308..975A}, which in MKS units is:
\begin{eqnarray}
f_{p}(T) &=& 10^{-35} T^{-2} \Bigl\{ 0.4 \exp \left[ -30 \left( \log_{10} T -4.6 \right)^{2} \right] \nonumber \\ 
&&+ 4 \exp \left[ -20 \left( \log_{10} T -4.9 \right)^{2} \right] \nonumber \\ 
&&+ 4.5 \exp \left[ -16 \left( \log_{10} T - 5.35 \right)^{2} \right] \nonumber \\
&&+ 2 \exp \left[ -4 \left( \log_{10} T -6.1 \right)^{2} \right] \Bigr\}.
\end{eqnarray}

To compare with the wave heating term, we overplot in Figure~\ref{fig:balance} the thermal conduction term and the cooling term in absolute values. We find that radiative cooling is the largest term in the hot, evacuated part of the thread, whereas wave heating dominates in the central, cool region. Near the ends of the thread, although  radiative cooling is still dominant, thermal conduction reaches its largest contribution. In the central part, thermal conduction can become more important than radiation because of the enhancing effect of the neutral conductivity.   The radiative cooling and the thermal conduction terms  depend on the temperature profile. When comparing the LTE and non-LTE cases we do not see important differences in the totally ionized region, where the temperature profiles are almost identical in  the LTE and non-LTE cases. We note that owing to the influence of the density profile, the maximum of the cooling function is not located at the ends of the thread, but at both sides of the densest region where the conditions are equivalent to those of prominence-corona transition region (PCTR). Because of thermal conduction, there is a flow of energy from the hot part of the thread towards the  cool central part.  We find that the thermal conduction term changes sign within the cool region at some specific locations. There, this term actually behaves as a heating term.

\begin{figure*}[!htb]
    \centering
    \includegraphics[width=9cm,height=6cm]{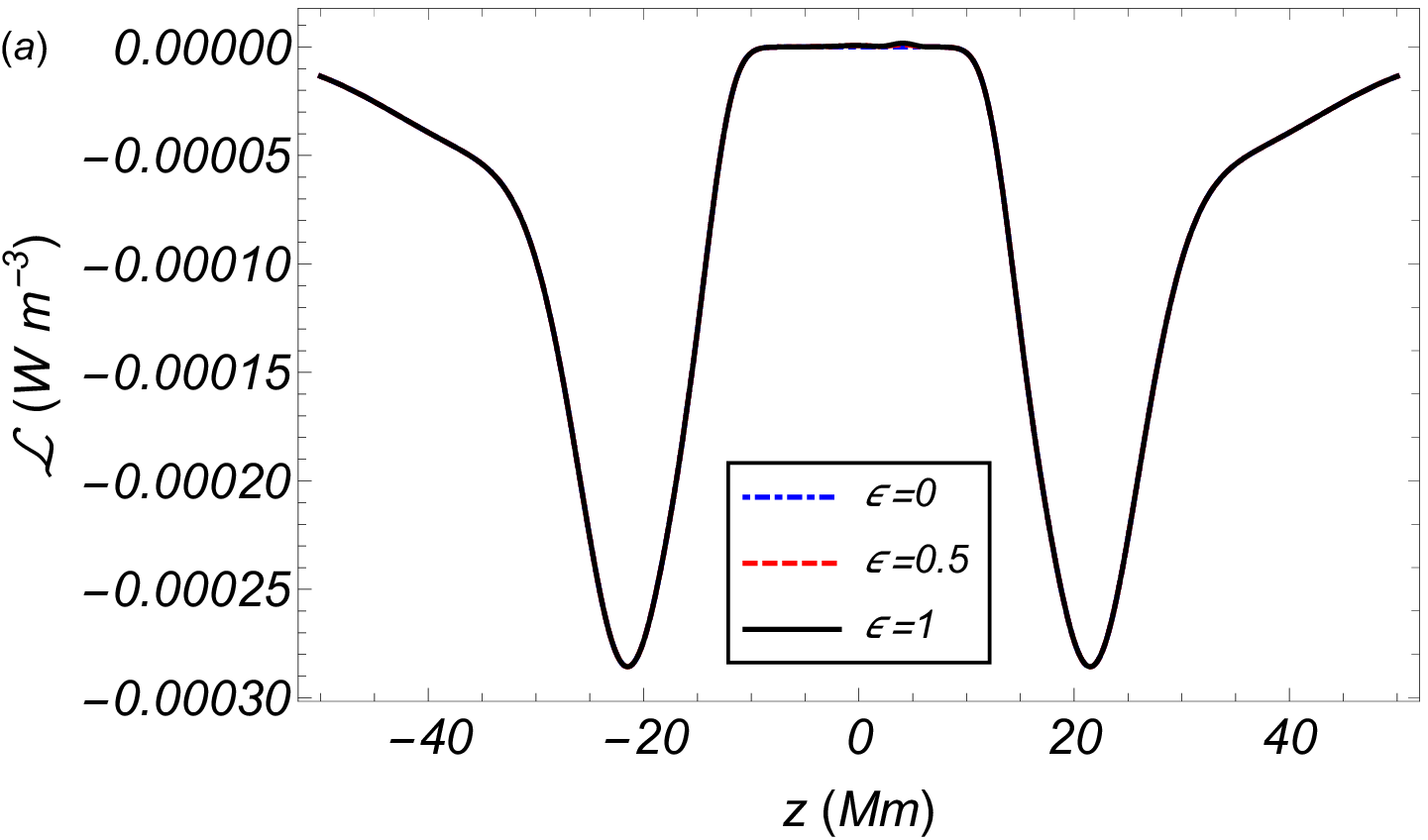} 
    \includegraphics[width=9cm,height=6cm]{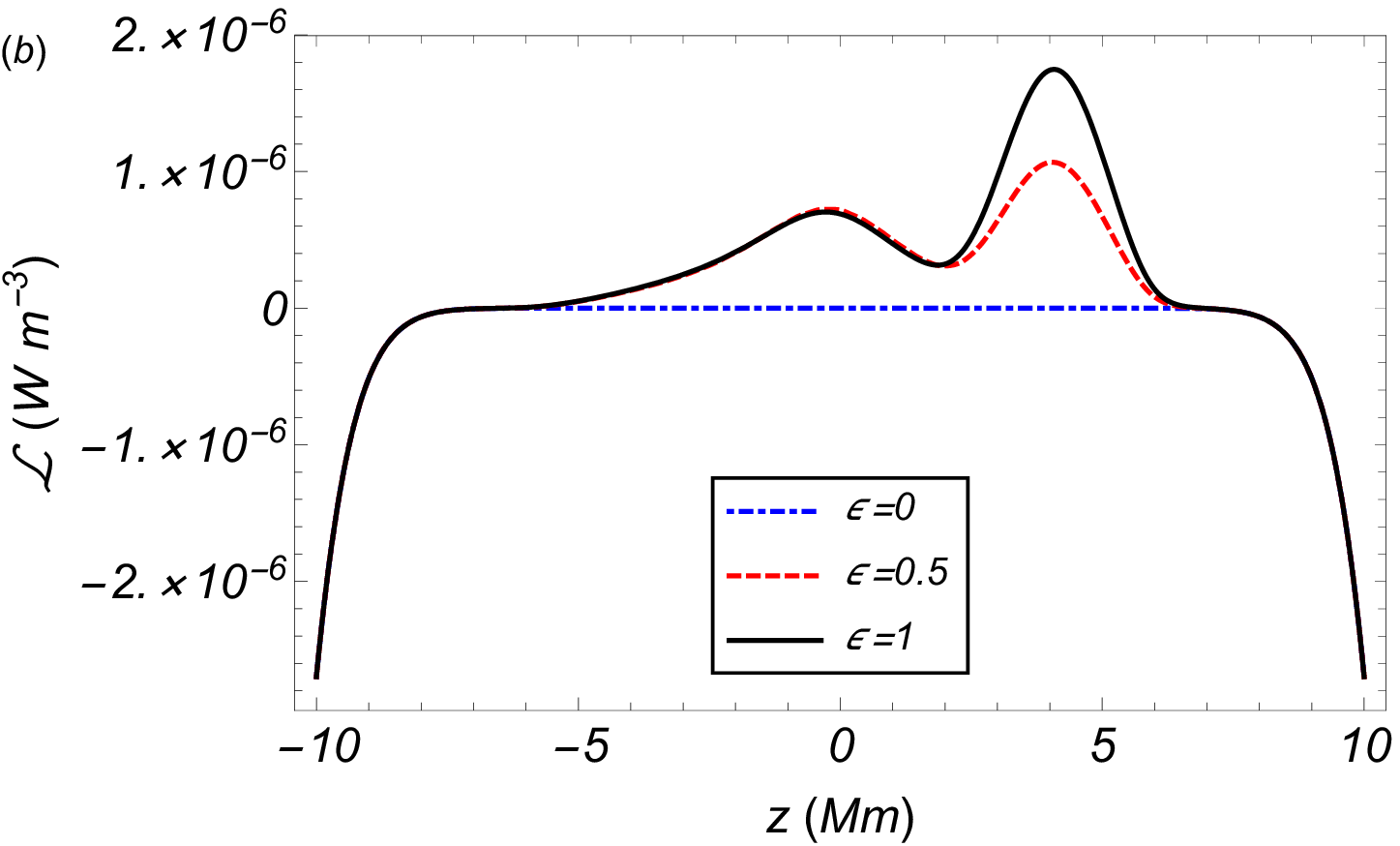} 
    \includegraphics[width=9cm,height=6cm]{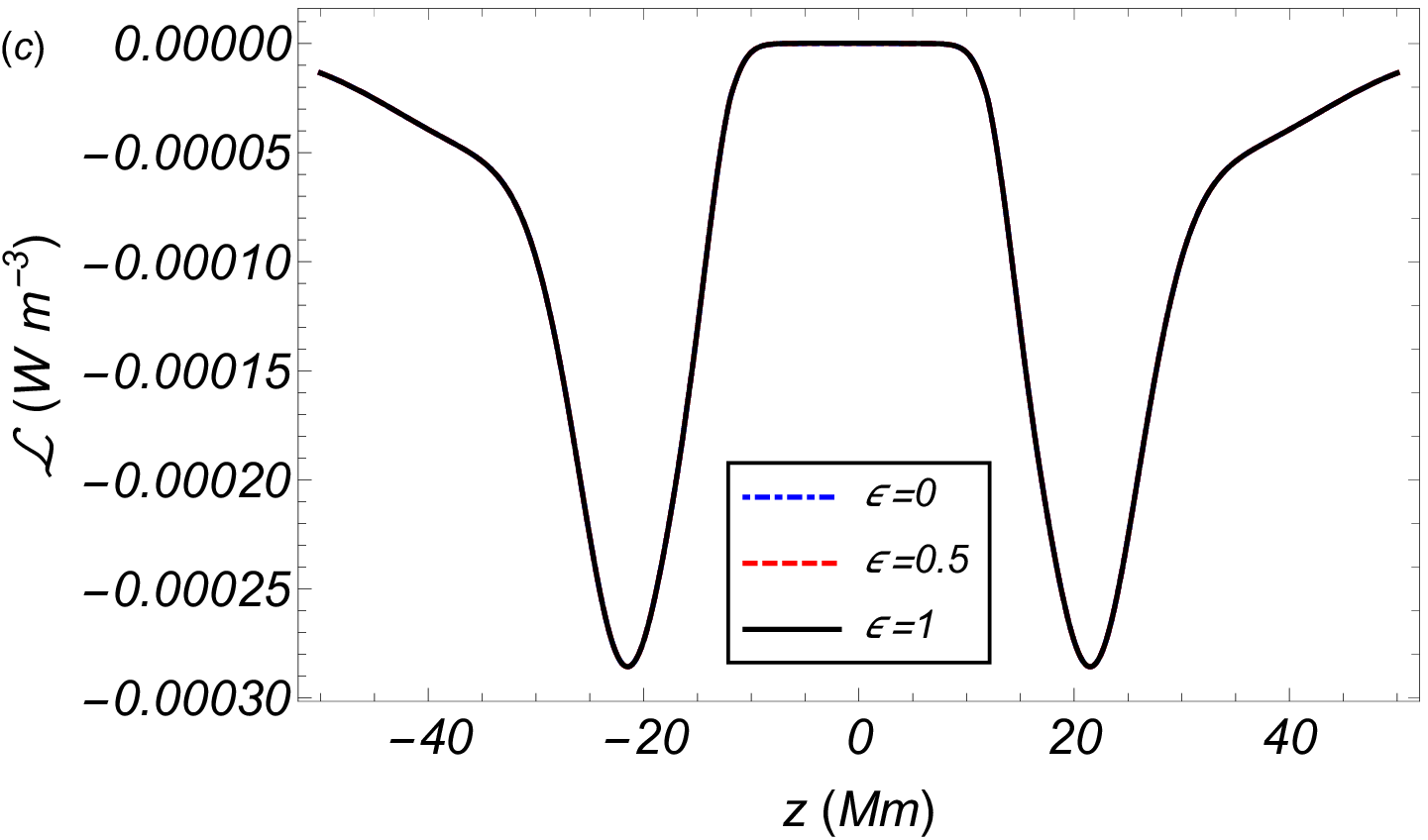} 
    \includegraphics[width=9cm,height=6cm]{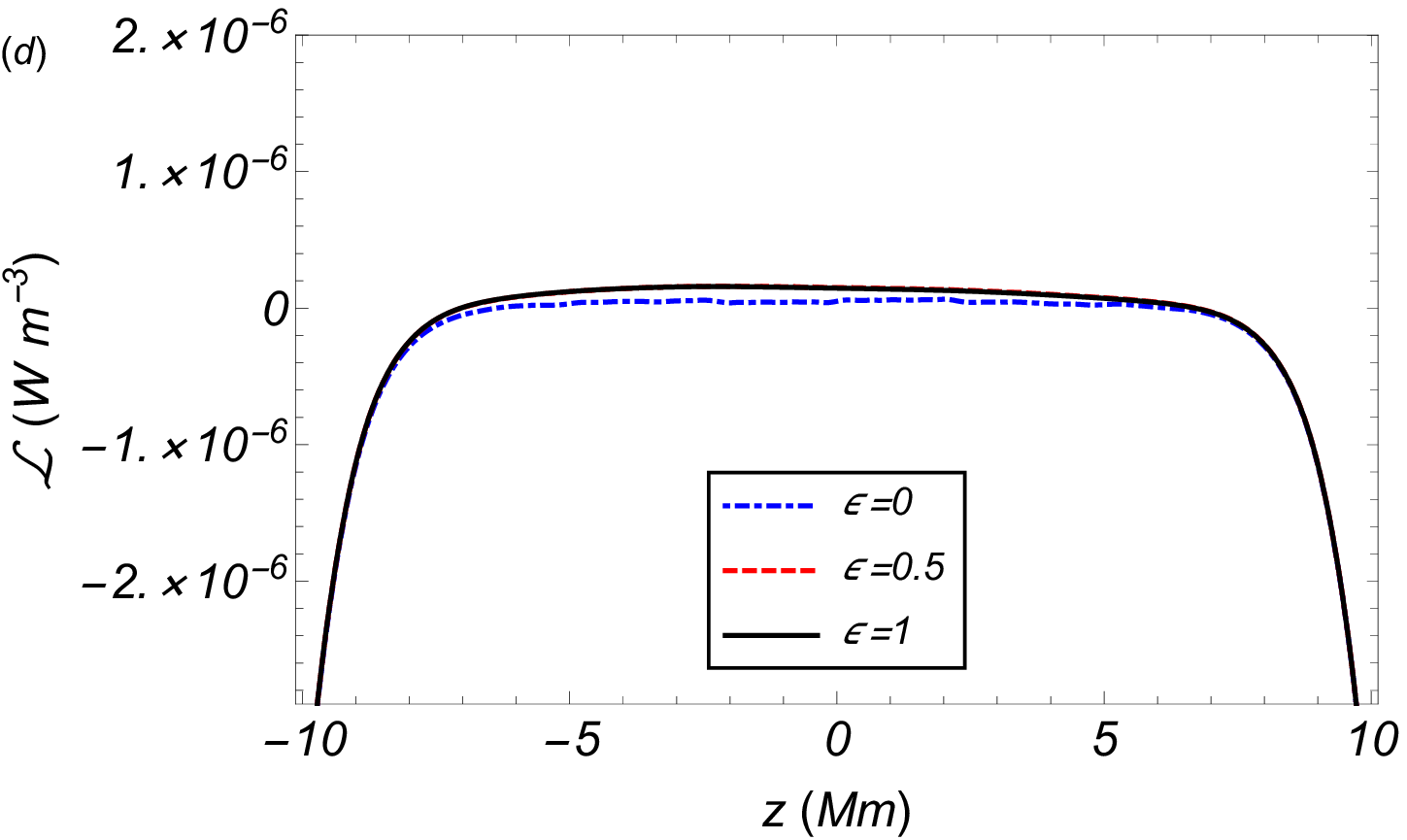} 
    \caption{Heat-loss function along the thread for the LTE method: \textbf{a)} complete domain and \textbf{b)} close-up view of the central part of the thread; \textbf{c)} and \textbf{d)} Same as \textbf{a)} and \textbf{b)} but for the non-LTE method.}
    \label{fig:heatloss}
\end{figure*}

In a situation of thermal equilibrium, $\mathcal{L} = 0$. To determine how far we are from thermal equilibrium, we plot the complete heat-loss function in Figure \ref{fig:heatloss}.  In the hot part of the thread, the heat-loss function has almost the same form for both LTE and non-LTE cases and for the three boundary conditions due to the dominance of  radiative cooling. The heat-loss function is negative along the whole thread (again due to the radiative cooling) except at the central, densest zone. The minimum value of the heat-loss function is $\mathcal{L} \approx - 3 \times 10^{-4}$~W~m$^{-3}$ and is located  at $z \approx \pm 20$~Mm, which corresponds to the location where  radiative cooling is most efficient. This is located within the PCTR of the thread. Since we have considered no  heating due to incident coronal radiation in this simple estimation, cooling cannot be compensated for in the PCTR and coronal part of the thread.  This is a logical and expected result.

Conversely, in the central, densest part of the thread, corresponding to conditions of prominence core, the wave heating function reaches its highest values, while there is much less cooling  than in the evacuated parts. As a consequence, the cooling can be compensated for. There is even a slight excess of heating around the thread center, so that $\mathcal{L}$ becomes positive in that region. In the LTE case, we find $\mathcal{L} \sim 10^{-6}$~W~m$^{-3}$, while in the non-LTE case, an even smaller value of $\mathcal{L} \sim 10^{-7}$~W~m$^{-3}$ is present. Therefore, in the thread core the energy balance situation is very close to that of thermal equilibrium, especially in the non-LTE case. The effect of the different boundary conditions is minor in this regard. We finally note that the heating rates calculated here depend on the injected energy flux by the driver. As explained in Section~\ref{sec:BC}, we considered an injected flux of $10^{-2}$~W~m$^{-2}$ that is consistent with previous results on the transmission of photospherically-driven Alfv\'en through the chromosphere \citep{soler2019}. If the injected flux is increased or decreased, the resulting heating rates would be larger or smaller. This would consequently affect the energy balance estimation.

\section{Concluding remarks}
\label{sec:conc}

In this exploratory paper, we study the prospects of Alfvén wave dissipation as a heating mechanism in thin threads of solar prominences. The present work builds upon the previous paper by \citet{soler2016}, where a more rudimentary prominence model was used. Here, we consider a 1D model for a thin prominence thread, which is still simple but incorporates several important ingredients that were missing in \citet{soler2016}. For instance, the model is fully non-uniform so that the density, temperature, ionization degree, etc. vary along the thread. In addition, a more consistent representation of the Alfvén wave driver is considered here, including an observationally-motivated spectral weighting function.

The magnetic field and velocity perturbations associated with the Alfvén waves have been obtained by solving the linearized MHD equations for a partially ionized plasma in the single-fluid approximation. Assuming an injected energy flux consistent with the studies of the photospherically-driven Alfv\'en waves that are transmitted through the chromosphere \citep[see][]{soler2019}, the obtained amplitudes for the Alfv\'enic perturbations in the thread are similar to those reported in the observation of prominence oscillations \citep{arregui2018}. Moreover, the perturbations amplitudes are much smaller than the background values, which justifies the linearization adopted at the beginning of the study. During the solution of the linearized MHD equations, we considered  three different boundary conditions, namely total reflection, partial reflection, and total transmission, at the right end of the thread. However, the results are very similar in the three cases, which highlights the robustness of the conclusions.


Ambipolar diffusion due to partial ionization of the plasma is a dissipation mechanism for Alfvén waves. The efficiency of this process depends on the ionization degree. An important part of the paper has been devoted to compare the results obtained when the temperature and ionization profiles are  computed considering either  LTE or non-LTE approximations for the hydrogen ionization. In general, the results are very similar in the two cases, although  a somewhat larger efficiency of wave dissipation is found in the LTE case. The reason for this result is that, for core prominence temperatures, the ionization degree computed in the LTE case is much lower than that computed in the non-LTE approximation. The consequence of this is  an enhanced efficiency of ambipolar diffusion.

Alfv\'en wave heating in the model is caused by Cowling's diffusion, that is, the joint effect of Ohm's and ambipolar diffusion. The computed Alfv\'en wave heating is largest at the thread center, where the density is highest and the temperature and ionization degree are lowest. At this spot, ambipolar diffusion is dominant. Conversely, in the hot coronal part of the model, the plasma is fully ionized and the Alfv\'en wave heating is negligible. In the fully ionized part, ambipolar diffusion is not present and the dissipation of wave energy is owing to Ohm's diffusion alone. For coronal conditions, the coefficient of Ohm's diffusion is so small that no appreciable wave heating is obtained.

In order to determine the relevance of wave heating, we estimated the energy balance by taking into account radiative cooling and thermal conduction in addition to wave heating. We assumed the cooling function after \citet{1986ApJ...308..975A} and the thermal conduction coefficient, including the contributions of electrons and neutrals. We found that in the coronal and PCTR parts, the model is far from being in thermal balance because of the predominance of cooling. We have not included a coronal heating function in this simple estimation, so that this result was expected because there is no mechanism in the corona that could counterbalance the lost of energy due to radiation. On the contrary, a situation of almost equilibrium is achieved in the central part of the thread, especially in the non-LTE case. There is even a slight excess of heating around the thread center. So, in the present case, the efficiency of wave heating in prominence conditions is found to be larger than that anticipated by \citet{soler2016} in an over-simplified model. This last result encourages us to further investigate  the role of wave heating in the prominence energy balance using more realistic models.

Another relevant conclusion that can be drawn is that the ad hoc density profile assumed in this work should evolve towards a new profile that would satisfy the energy balance if the background model was allowed to vary with time. Then, a new wave heating function could be computed for the new density profile, which  would in turn drive the system towards a different equilibrium.  This process could iteratively be  repeated all over again until convergence towards a self-consistent model would be achieved.  The process would be very similar to that of \citet{ofman1998} in the case of heating by resonant absorption. This interesting study would be a natural follow-up to the present paper and may be tackled in a forthcoming work.


\begin{acknowledgement}
We acknowledge the support from grant AYA2017-85465-P (MINECO/AEI/FEDER, UE).
\end{acknowledgement}

\bibpunct{(}{)}{;}{a}{}{,} 
\bibliographystyle{aa} 
\bibliography{refs} 

\begin{appendix}

\section{Standing modes}
\label{app}

Here we briefly study the standing modes of the thread, which are used to explain the existence of the resonances with the propagating waves. We use Equations~(\ref{eq:moment}) and (\ref{eq:induct}) with the temporal dependence specified as $\exp \left( - i \omega t  \right)$. The equations become
\begin{eqnarray}
\omega v_{\perp} &=& \frac{iB}{\rho \mu_{0}}  \frac{\partial B_{\perp}}{\partial z}, \label{eq:momd} \\
\omega B_{\perp} &=& iB\frac{\partial v_{\perp}}{\partial z} + i \frac{\partial}{\partial z} \left( \eta_{C}\frac{\partial B_{\perp}}{\partial z} \right). \label{eq:indd}
\end{eqnarray}
Equations (\ref{eq:momd}) and (\ref{eq:indd}) form an eigenvalue problem, where $\omega$ is the eigenvalue and $B_{\perp}$ and $v_{\perp}$ are the eigenfunctions. The eigenvalue problem is numerically solved after imposing the boundary conditions
\begin{eqnarray}
    v_{\perp} &=& 0 \qquad \textrm{at} \qquad z = \pm \frac{L}{2}, \\
    \frac{\partial B_{\perp}}{\partial z} &=& 0 \qquad \textrm{at} \qquad z = \pm \frac{L}{2}.
\end{eqnarray}
The eigenfunctions and eigenfrequencies are obtained for both LTE and non-LTE cases.

\begin{figure}[!htb]
    \centering
    \includegraphics[width=9cm,height=6cm]{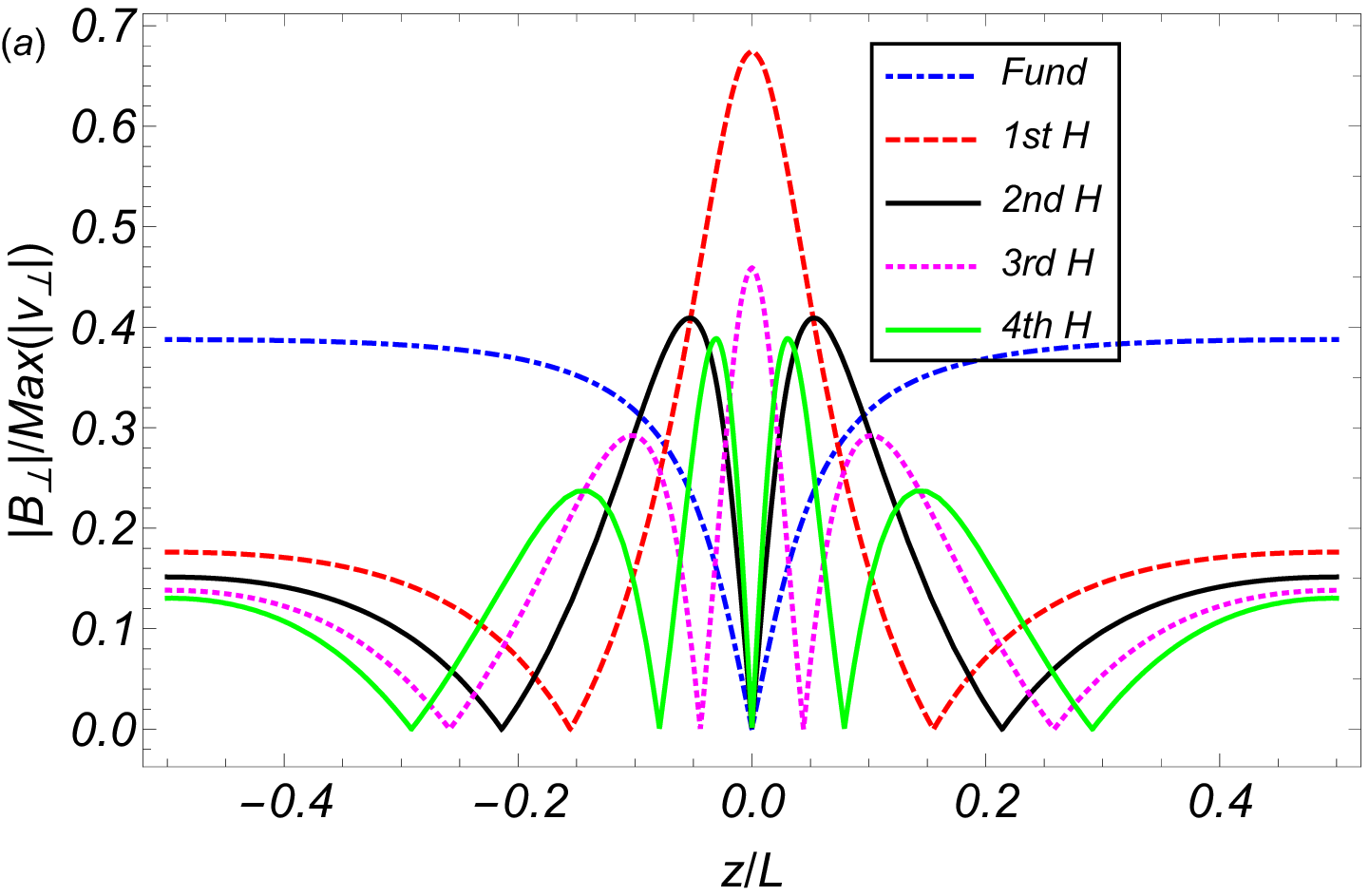} 
    \includegraphics[width=9cm,height=6cm]{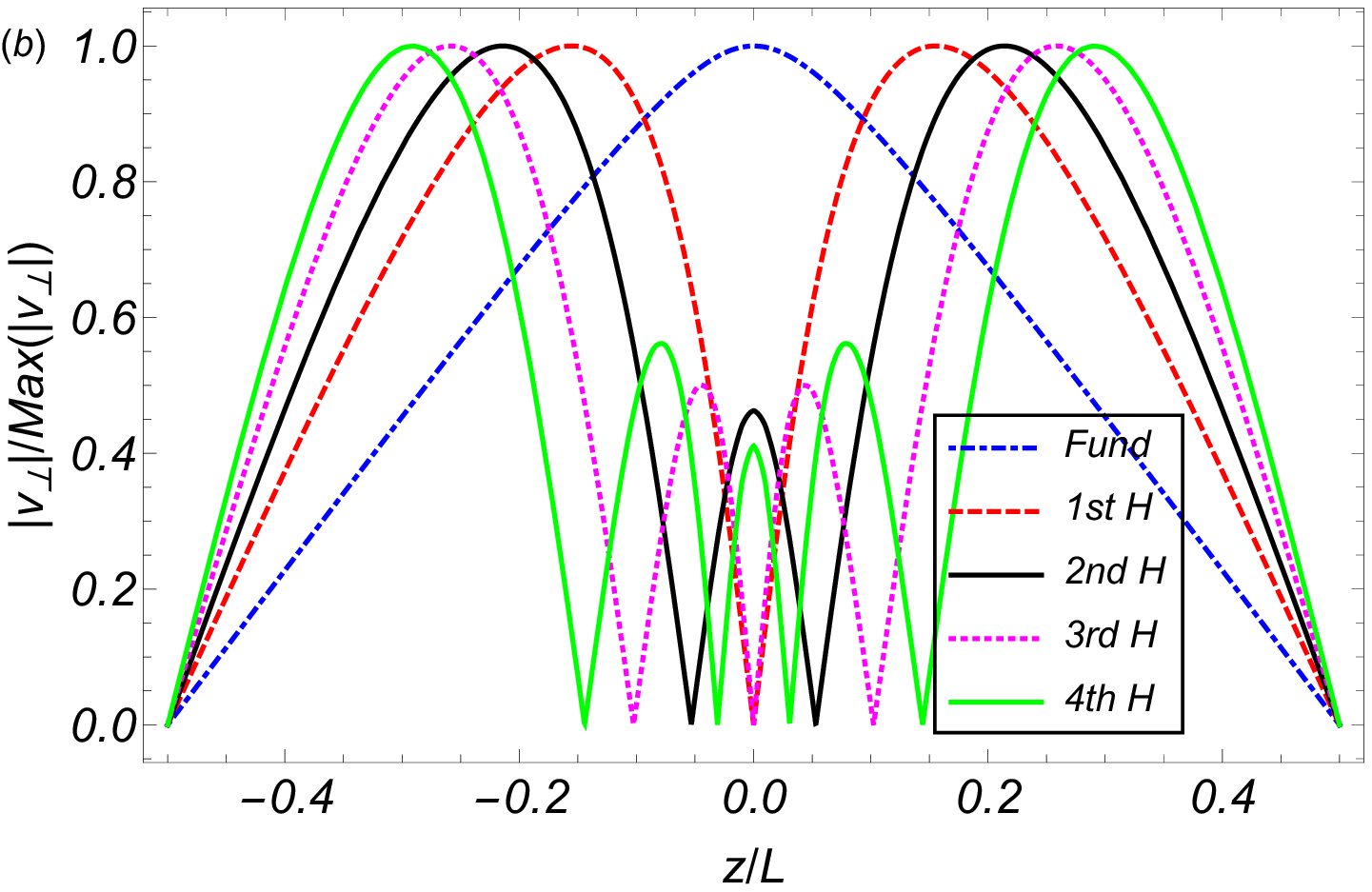} 
    \caption{Absolute value of the eigenfunctions of the fundamental mode and the first 4 harmonics of standing modes in the non-LTE case: \textbf{a)} magnetic field perturbation and \textbf{b)} velocity perturbation. The eigenfunctions are normalized to the maximum of the velocity perturbation in all cases. The eigenfunctions in the LTE case are indistinguishable from those displayed here.}
    \label{fig:eigenfunction}
\end{figure}

 Figure~\ref{fig:eigenfunction} shows the eigenfunctions of the fundamental mode and the first four harmonics in the non-LTE case. Both perturbations have a clear harmonic-like behavior with a spatially-dependent amplitude that is determined by the Alfv\'en speed profile along the thread. The amplitude of the magnetic field perturbations are larger in the densest part of the thread. On the contrary, except for the fundamental mode the velocity perturbations tend to have their maximums  located further away from the center, especially as the mode number increases. The shapes of the eigenfunctions in the LTE case are almost identical to those of the non-LTE case and are not shown here for simplicity.

Following \citet{1992A&A...261..625J} and \citet{1993ApJ...409..809O}, the standing modes can be classified according to the location of the maximum of the velocity perturbation. If the maximum is in the dense prominence region, the mode is called an internal mode. If the maximum is in the evacuated coronal part, the mode is called an external mode. If the velocity is large in both regions, the mode is called a hybrid or string mode. According to Figure \ref{fig:eigenfunction}b, the only mode that could be considered as an internal or hybrid mode is the fundamental one. All overtones should be classified as external modes. However, we note that the classification of \citet{1992A&A...261..625J} and \citet{1993ApJ...409..809O} applies to a slab model  with a piecewise constant density, while here we use a continuous density profile.

\begin{figure}[!htb]
    \centering
    \includegraphics[width=9cm,height=6cm]{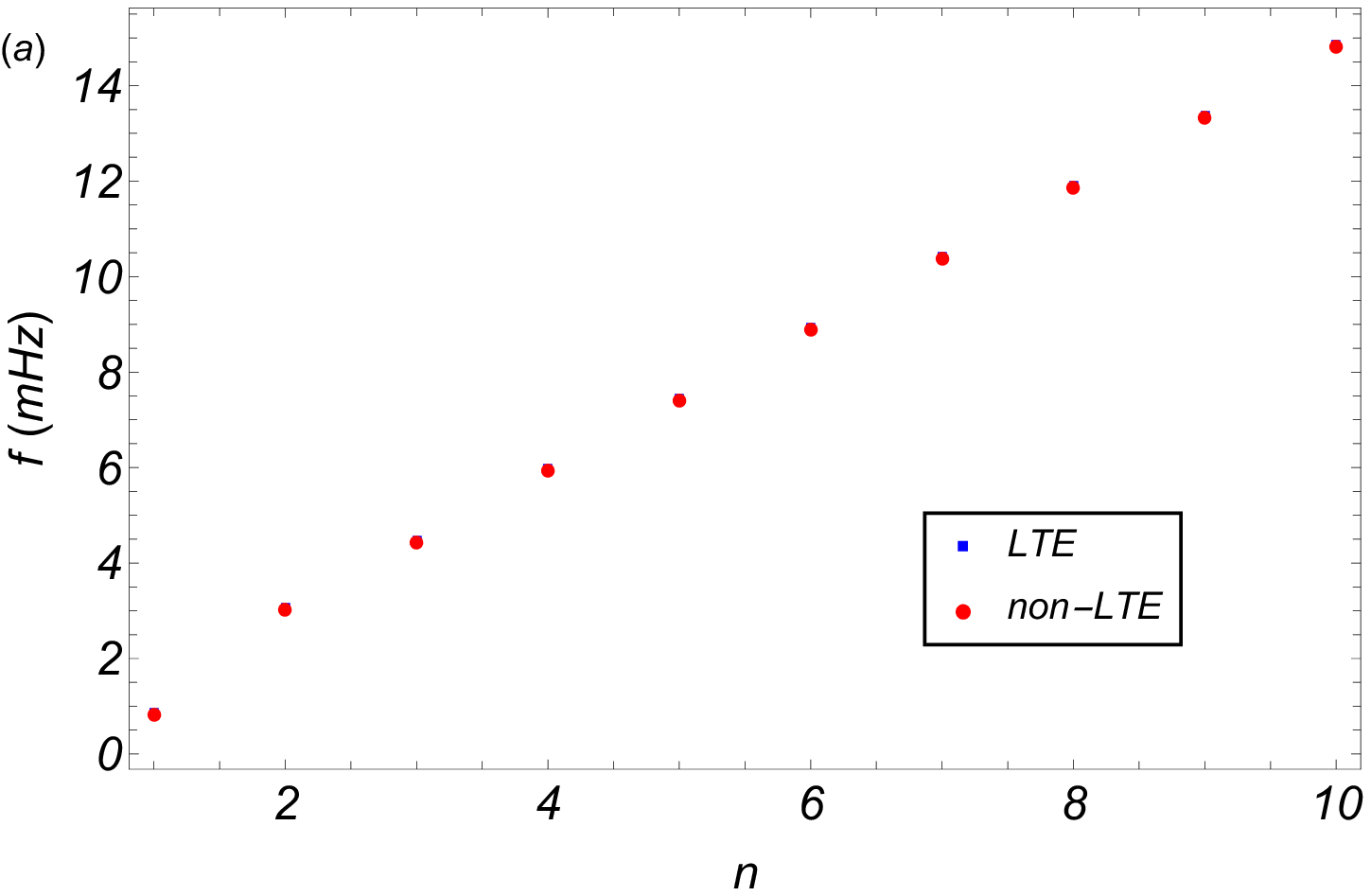} 
    \includegraphics[width=9cm,height=6cm]{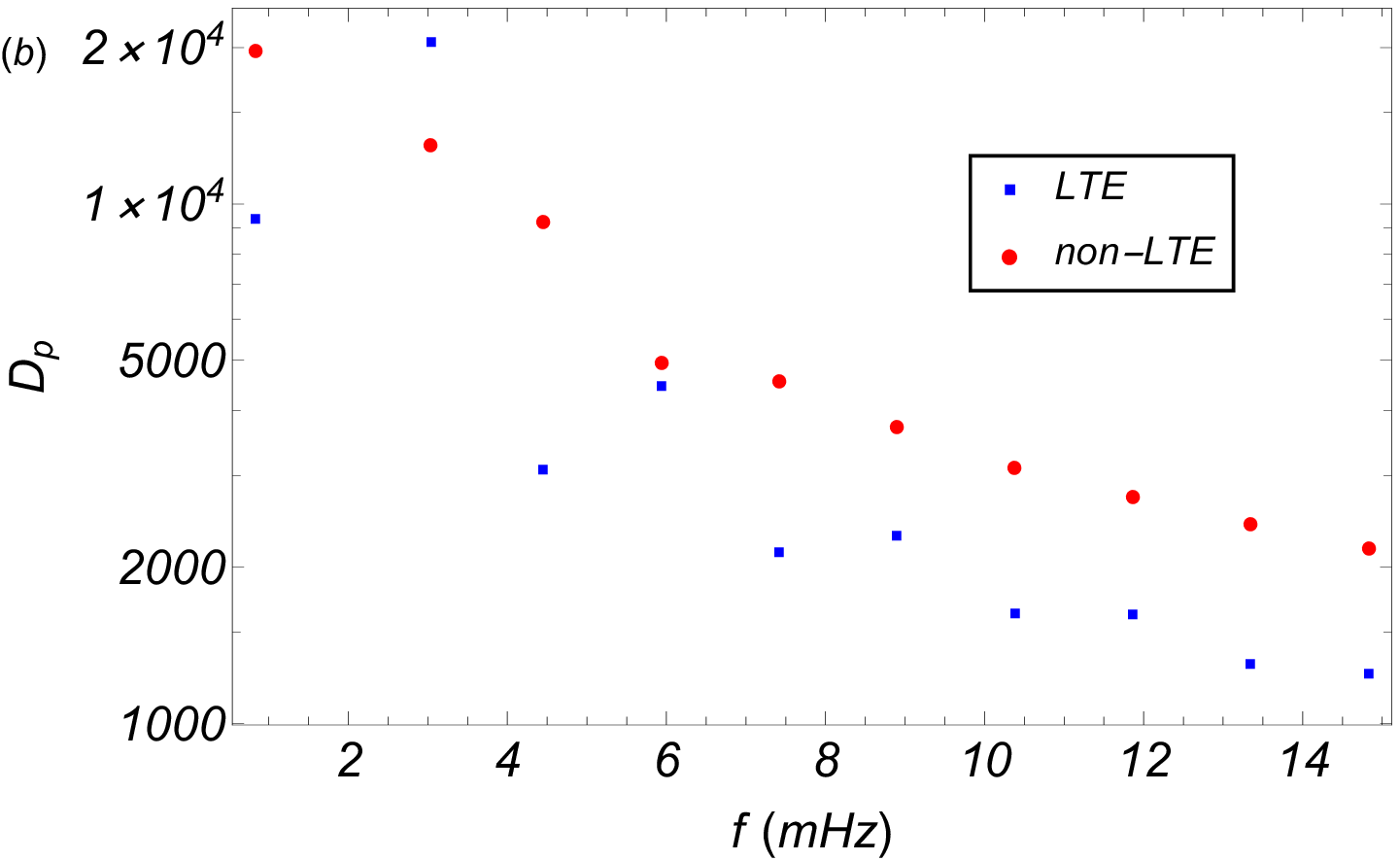} 
    \caption{\textbf{a)} Real part of the eigenfrequency as function of the mode number for the first 10 modes. \textbf{b)} Damping per period as a function of the eigenfrequency for the same modes as in \textbf{a)}.}
    \label{fig:eigenfrequency}
\end{figure}

The eigenfrequencies for the first 10 modes are shown in Figure \ref{fig:eigenfrequency}a, where $f=\omega_{\rm R}/2\pi$ with $\omega_{\rm R}$ the real part of the eigenfrequency. The eigenfrequencies show an almost linear dependence with the mode number and almost the same result is obtained for both LTE and non-LTE cases. This can be explained by the lack of influence of  Cowling's diffusion on the real part of the eigenfrequencies. The different profiles of Cowling's coefficient is the distict feature between both cases (Figure~\ref{fig:magnitudes}f). 

The damping per period measures the efficiency of the damping  and is defined as $D_{p}=\omega_{\rm R}/2\pi\omega_{\rm I}$, where $\omega_{\rm I}$ is the imaginary part of the eigenfrequency.  Figure \ref{fig:eigenfrequency}b shows the damping per period of the first 10 modes as a function of their corresponding $f$. In both LTE and non-LTE cases, the damping per period has extremely large values. The damping per period decreases as the eigenfrequency increases, but we should consider very high harmonics for the damping to be significant. The damping per period is larger in the non-LTE case than in the LTE case, which is again related to the relative values of the Cowling's coefficient in the two cases. 

The result that the damping of the standing modes is very inefficient implies that the heating associated with the dissipation of their energy  is unimportant for all practical purposes. If standing modes are resonantly excited by the propagating waves, as the results of the paper evidence, part of the energy of the propagating waves should be transferred to the standing modes. However, since the standing modes are practically undamped, their contribution to the plasma heating would be negligible. This explains why the heating rates computed in the paper have relative minimums at the resonance frequencies.

\end{appendix}
\end{document}